\documentclass{article}

\usepackage{emulateapj}
\usepackage{onecolfloat}
\usepackage{graphicx} 
\usepackage{fancyheadings} 

\def\msol{M_\odot}

\def\delv{\Delta v}

\def\lya{Ly$\alpha$ }
\def\ndmp{31 }

\def\kms{km~s$^{-1}$ }
\def\cm#1{\, {\rm cm^{#1}}}
\def\N#1{{N({\rm #1})}}
\def\f#1{{f_{\rm #1}}}

\def\sci#1{{\rm \; \times \; 10^{#1}}}
\def\ltk{\left [ \,}
\def\ltp{\left ( \,}

\def\rtk{\, \right  ] }
\def\rtp{\, \right  ) }

\def\ohf{{1 \over 2}}

\def\rhf{{3 \over 2}}

\def\perd{\;\;\; .}
\def\cmma{\;\;\; ,}

\def\Nperp{N_{\perp} (0)}
\def\intl{\int\limits}

\def\sphr{\sqrt{R^2 + Z^2}}
\def\vrot{v_{rot}}
\def\btau{\bar\tau (v_{pk}) / \sigma (\bar \tau)}
\def\Ipk{I(v_{pk})/I_{c}}
\def\mkms{{\rm \; km\;s^{-1}}}
\newcommand{\tskip}{\tablevspace{3pt}}

\begin{document}

\twocolumn[%
\submitted{ApJ: Accepted May 21, 1998}

\title{PROTOGALACTIC DISK MODELS OF DAMPED \lya KINEMATICS}

\author{ JASON X. PROCHASKA\altaffilmark{1} 
\& ARTHUR M. WOLFE\altaffilmark{1}\\
Department of Physics, and Center for Astrophysics and Space Sciences; \\
University of California, San
Diego; \\
C--0424; La Jolla; CA 92093\\}

\begin{abstract} 

We present new observational results on the kinematics of the
damped \lya systems.  Our full sample is now comprised of 31 low-ion
profiles and exhibits similar characteristics to the sample from Paper I.
The primary exception is that the new distribution of velocity
widths includes values out to a maximum of nearly 300~\kms, 
$\approx$ 100~\kms greater than the previous maximum.  
These high velocity width systems will significantly leverage
models introduced to explain the damped \lya systems.
Comparing the characteristics from low-redshift and high-redshift
 sub-samples, we find no evidence for significant
evolution in the kinematic properties of protogalaxies
from $z = 2.0 - 3.3$.  

The new observations give
greater statistical significance to the main conclusions of our
first paper.  
In particular, those models inconsistent with the damped 
\lya observations
in Paper I are ruled out at even higher levels of confidence.
At the same time, the observations are consistent with a population
of rapidly rotating, thick disks (the TRD model) at high redshift,
as predicted by cosmologies with early structure formation.

Buoyed by the success of the TRD model, we investigate
it more closely by considering more
realistic disk properties.  Our goal is to demonstrate the statistical
power of the damped \lya observations by investigating the robustness
of the TRD model.  
In particular, we study the effects of warping,
realistic rotation curves, and photoionization on the kinematics
of disks in the TRD model.
The principal results are: (1) disk warping has only
minimal effect
on the kinematic results, primarily influencing the effective
disk thickness, (2) the TRD model is robust to more realistic
rotation curves; we point out, however, that the rotation curve derived
from centrifugal equilibrium with
HI gas alone does not yield acceptable results,
rather flat rotation curves such as those generated by dark matter
halos are required,
and (3) the effects of
photoionization require thicker disks 
to give consistent velocity width distributions. 
\end{abstract}

\keywords{cosmology---galaxies: evolution---galaxies: 
quasars---absorption lines}]

\altaffiltext{1}{Visiting Astronomer, W.M. Keck Telescope.
The Keck Observatory is a joint facility of the University
of California and the California Institute of Technology.}

\pagestyle{fancyplain}
\lhead[\fancyplain{}{\thepage}]{\fancyplain{}{PROCHASKA \& WOLFE}}
\rhead[\fancyplain{}{PROTOGALACTIC DISK MODELS}]{\fancyplain{}{\thepage}}
\setlength{\headrulewidth=0pt}
\cfoot{}

\section{INTRODUCTION}

This paper marks the third in a series of papers on the kinematics
of the damped \lya protogalaxies.  These HI gas layers observed
along sightlines to distant QSO's are widely believed to be the
gaseous progenitors of modern galaxies (\cite{wol95b}).
Hence an examination
of damped systems at high redshift provides insight into
the process of galaxy formation.  For instance, identifying the
physical nature of these systems may 
distinguish between the monolithic collapse model
(\cite{egg62}) and the hierarchical scenario favored
by standard cosmogony.  In our first paper (\cite{pro97b},
hereafter PW), we demonstrated that the kinematics of damped \lya systems
at high redshift are consistent with these systems being thick,
rapidly rotating disks; it is a description not unlike that predicted by
monolithic collapse formation scenarios.  At the same time, we found
damped \lya systems cannot be
simple exponential disks in a cluster normalized Standard Cold
Dark Matter cosmology (e.g.\ \cite{kau96}).
Subsequently, Jedamzik and Prochaska (1998) tightened this conclusion
by considering a range of disk characteristics and CDM normalizations. 
They found that only a finely 
tuned disk model within the framework of
CDM could be made marginally consistent with
the damped \lya observations.
Recently, Haehnelt et al.\ (1997)
have offered an alternative model for damped systems as 
gaseous protogalactic clumps undergoing infall within 
dark matter halos which may be consistent with the kinematic characteristics
of the damped \lya systems.  Such a description lends itself naturally
to the hierarchical cosmologies where merging plays a vital role.
A future paper will address this model in greater detail. 
For the present work, we will
focus on the interpretation of damped \lya systems as thick
rotating disks at high redshift. 

In PW, we analyzed the low-ion profiles from 17 damped \lya systems
and compared their kinematic
characteristics with those of simulated profiles 
derived from several physical
models.  Of the models tested, we found the thick
rotating disk model
to be the only model consistent with the observations.
The basic assumptions of the model
are a flat rotation curve and an exponential
gas distribution, both chosen to roughly
correspond with the observations of local spiral galaxies. 
In this paper we present low-ion profiles for 14 additional
damped systems.  Therefore, our full kinematic sample consists of
31 low-ion profiles.  We interpret the sample with disk models 
containing
more physically realistic characteristics, e.g. disk warping and 
photoionization.  We have two primary goals in mind:
(1) to test the robustness of the interpretation of damped \lya
systems as disks and (2) to determine the effects on our conclusions
regarding the thickness and rotation speed of these disks.

In $\S 2$ we review the terminology and methodology introduced
in PW.  The new data are presented and tested against the models
of PW in $\S 3$.  We investigate the effects of more realistic
disk properties in $\S 4$ and in $\S 5$ we present a summary.

\section{A REVIEW}

Our strategy is to compare the kinematic
characteristics of low-ion profiles in damped \lya systems
with the simulated profiles derived from various physical models.
In this manner, then, we investigate the agreement between the
models and the damped systems.
In this section we review the methodology
and the thick rotating disk (TRD) model presented in PW.

To characterize the kinematics of the damped systems we 
focus on low-ion profiles such as SiII 1808.
After normalizing them to unit continuum strength, we create
a binned apparent optical depth array, $\bar \tau(v)$, where
$\bar\tau(v) = <\ln[I_c/I(v)]>$, $I(v)$ is the intensity at velocity $v$,
$I_c$ is the continuum intensity, 
and the average is taken
over $\approx$~two resolution elements centered at $v$.
We focus on $\bar\tau (v)$ rather than $I(v)$ to minimize the effects
of saturation, and we smooth in
optical depth to both limit the effects of Poisson noise and
focus on features as opposed to individual components.
Each profile is characterized by its
signal-to-noise ratio (SNR), instrumental resolution, and
peak optical depth ratio,
$\btau$,  with $v_{pk}$ the position of the maximum of $\bar\tau(v)$
in velocity space.

In PW, the only model considered which is consistent with the damped
\lya kinematic characteristics is the TRD model.
For this model, we assume:

\begin{itemize}

\item An exponential gas distribution

\begin{equation}
n(R,Z) = n_0 \exp \ltk - \, {R \over R_d} - {|Z| \over h}  \rtk \;\;\; ,
\label{TRDvol}
\end{equation}

\noindent where $R$ and $Z$ are
cylindrical radius and vertical displacement from midplane, 
$n_0$ is the central gas density,
$R_{d}$ is the radial scale length, and $h$ is the vertical
scale height.  Note $n_0$ is related to the perpendicular column density
at $R=0$ by $\Nperp = 2 n_0 h$.

\item A flat rotation curve parameterized by the rotation speed,
$v_{rot}$  

\item A random Gaussian velocity field parameterized
by a 1-d velocity dispersion, $\sigma_{cc}$.
\end{itemize}

For the model systems considered here,
we derive simulated profiles with identical SNR,
instrumental resolution, and peak optical depth as the empirical data set.  
We employ the same Monte Carlo techniques as those used
in $\S3.2$ of PW.  However, we now consider \ndmp
damped \lya systems which includes the 17 systems used in PW.  
To sufficiently
sample the model distributions requires $\approx 10000$
sightlines, thus we perform \ndmp
runs of 333 sightlines for every simulation.
The SNR, $\btau$, and instrumental 
resolution of the simulated profiles for each
set of 333 sightlines is varied accordingly.  

We quantify the kinematic characteristics of the 
profiles with 4 test statistics:

\begin{itemize}
\item  Velocity Interval Test -- The width of the profiles in 
velocity space, $(\f{\delv})$ or $\delv$.  Formally, we calculate
the total optical depth, $\tau_{tot}$ of the profile in velocity space
and define $\delv$ to be the velocity width encompassing $0.9 \, \tau_{tot}$
by trimming $0.05 \, \tau_{tot}$ off each edge of the profile.

\item Mean-Median Test -- Measures the asymmetry of the profile

\begin{equation}
f_{mm} = {| v_{mean} - v_{median} | \over (\f{\delv} / 2)} \perd
\end{equation}

\item Edge-Leading Test -- Designates how edge-leading the peak of
the profile is

\begin{equation}
f_{edg} = {| v_{pk} - v_{mean} | \over ( \f{\delv} / 2)} \perd
\end{equation}

\item Two Peak Test -- Designates how edge-leading the 2nd peak of
the profile is

\begin{equation}
f_{2pk} = {\pm}\frac{\vert v_{2pk}- v_{mean} \vert} {(\f{\delv}/2)} \perd
\end{equation}

\end{itemize}

\noindent In the above equations, $v_{pk}$ and $v_{2pk}$ indicate the position 
of the peak and second strongest peak of the $\bar\tau(v)$ profile, and 
$v_{mean}$ and $v_{median}$ are the mean and median of the 
$\bar\tau(v)$ profile.  We perform the statistical tests on the damped
\lya and simulated profiles in the same manner and compare the
resulting distributions with the two-sided Kolmogorov-Smirnov (KS) Test.

\section{DAMPED LYA PROFILES}

\subsection{DATA SAMPLES AND SELECTION CRITERIA}

The full kinematic data sample, designated Sample B,
comprises 17 damped \lya systems from PW (Sample A)
and 14 new systems. 
Table~\ref{qsotab} contains the journal of 
the 14 new observations.
In Table~\ref{dlatab} we list the following properties of the 14 new
profiles: absorption redshift (column 2), log of $\N{HI}$ (column 3),
Fe abundance, i.e., $\log({\rm Fe/H}) - \log({\rm Fe/H})_\odot$ (column 4),
Zn abundance (column 5), the low-ion transition 
(column 6), and $\btau$ (column 7).
From each new damped \lya system we select a 
low-ion transition satisfying strict selection criteria.  
However, in order to include the highest
redshift systems, we relax some of
the selection criteria established in PW.  This is because
(1) the data have poorer SNR
and (2) the high $z$ systems have lower metallicity.
Thus, even strong transitions (e.g.\ Fe II 1608) are difficult to measure;
for

\begin{table*}[ht] 
\caption{\centerline{
{\sc QSO and Observational Data} \smallskip} \label{qsotab}}
\begin{center}
\begin{tabular}{llclccc}
\tableline
\tableline \tskip
QSO & Date
& Exp
& $z_{em}$
& Res & SNR & Data \\
& & (s) & & (km/s)& & set \\
\tableline \tskip
Q0019$-$01 & F96 & 35000 & 4.528 & 7.5 & 18 & W\tablenotemark{a} \nl
Q0347$-$38 & F96 & 12600 & 3.23  & 7.5 & 33 & W \nl
Q0951$-$04 & S97 & 30600 & 4.369 & 7.5 & 13 & W \nl
Q1005+36   & S97 & 3600  & 3.17  & 7.5 & 12 & W\tablenotemark{b} \nl
Q1346$-$03 & S97 & 31000 & 3.992 & 7.5 & 29 & W \nl
Q1759+40   & F96 & 10400 & 3.05  & 6.6 & 33 & W \nl
Q2348$-$14 & F96 & 9000  & 2.940 & 7.5 & 41 & W \nl
\nl
Q0930+28   & S97 & 17000 & 3.436 & 6.6 & 26 & SL\tablenotemark{c}\nl
Q1104$-$18 & S97 & 19500 & 2.32  & 6.6 & 40 & SL \nl
Q1850+40   & S97 & 11880 & 2.120 & 6.6 & 16 & SL \nl
Q2233+13   & S97 & 15000 & 3.274 & 6.6 & 16 & SL \nl
Q2343+34   & S97 & 40000 & 2.58  & 6.6 & 35 & SL \nl
Q2344+12   & S97 & 15000 & 2.763 & 6.6 & 29 & SL \nl
\tskip \tableline
\end{tabular}
\end{center}
\centerline{$^a$W - Data acquired by Wolfe \& Prochaska}
\centerline{$^b$QSO coordinates kindly provided by Becker (1998)}
\centerline{$^c$SL - Data kindly provided by W. L. W. Sargent, Limin Lu,
and collaborators}
\end{table*}

\begin{table*}[hb]
\caption{\label{dlatab}}
\begin{center}
{\sc Additional Sample of DLA Systems \smallskip}
\begin{tabular}{lcccccc}
\tableline
\tableline \tskip
QSO & $z_{abs}$ & $\log [\N{HI}]$ & [Fe/H] & [Zn/H] &
Transition & $\bar\tau(v_{pk}/\sigma_{\bar\tau})$ \\
\tableline \tskip
Q0019$-$01  & 3.439 & 20.9 & $> - 1.7$\tablenotemark{a} & N/A & FeII 1608 & 
\tablenotemark{b}  \nl
Q0347$-$38  & 3.025 & 20.8 & $- 1.8$  & N/A      & FeII 1608 & 33 \nl
Q0951$-$04A & 3.859 & 20.6 & $-2.04$  & N/A      & SiII 1526 & 
\tablenotemark{b} \nl
Q0951$-$04B & 4.203 & 20.4 & $-2.94$  & N/A      & SiII 1190 & 10 \nl
Q1005+36    & 2.799 & 20.6 & $-2.0$   & N/A      & FeII 1608 & 13 \nl
Q1346$-$03  & 3.736 & 20.7 & $< -1.7$ & N/A      & SiII 1304 & 29 \nl
Q1759+40    & 2.625 & 20.8 & $-1.27$  & N/A      & SiII 1808 & 41 \nl
Q2348$-$14  & 2.279 & 20.6 & $-2.35$  & $<-1.19$ & FeII 1608 & 40 \nl
\nl
Q0930+28    & 3.235 & 20.3 & $-2.03$  & N/A      & FeII 1608 & 31 \nl
Q1104$-$18  & 1.661 & 20.8 & $-1.40$  & $-0.865$ & SiII 1808 & 44 \nl
Q1850+40    & 1.990 & 21.4 & $-1.33$  & $-0.70$  & ZnII 2026 & 16 \nl
Q2233+13    & 3.149 & 20.0 & $-1.5$   & N/A      & FeII 1608 & 19 \nl
Q2343+34    & 2.430 & 20.3 & $-1.20$  & N/A      & SiII 1808 & 25 \nl
Q2344+12    & 2.538 & 20.4 & $-2.0$   & N/A      & AlII 1670 & 36 \nl
\tskip \tableline
\end{tabular}
\end{center}
\centerline{$^a$All limits are 3$\sigma$ limits}
\centerline{$^b$Profile is saturated at the peak}
\end{table*}

\clearpage

\begin{figure*}
\begin{center}
\includegraphics[height=8.0in, width=6.0in]{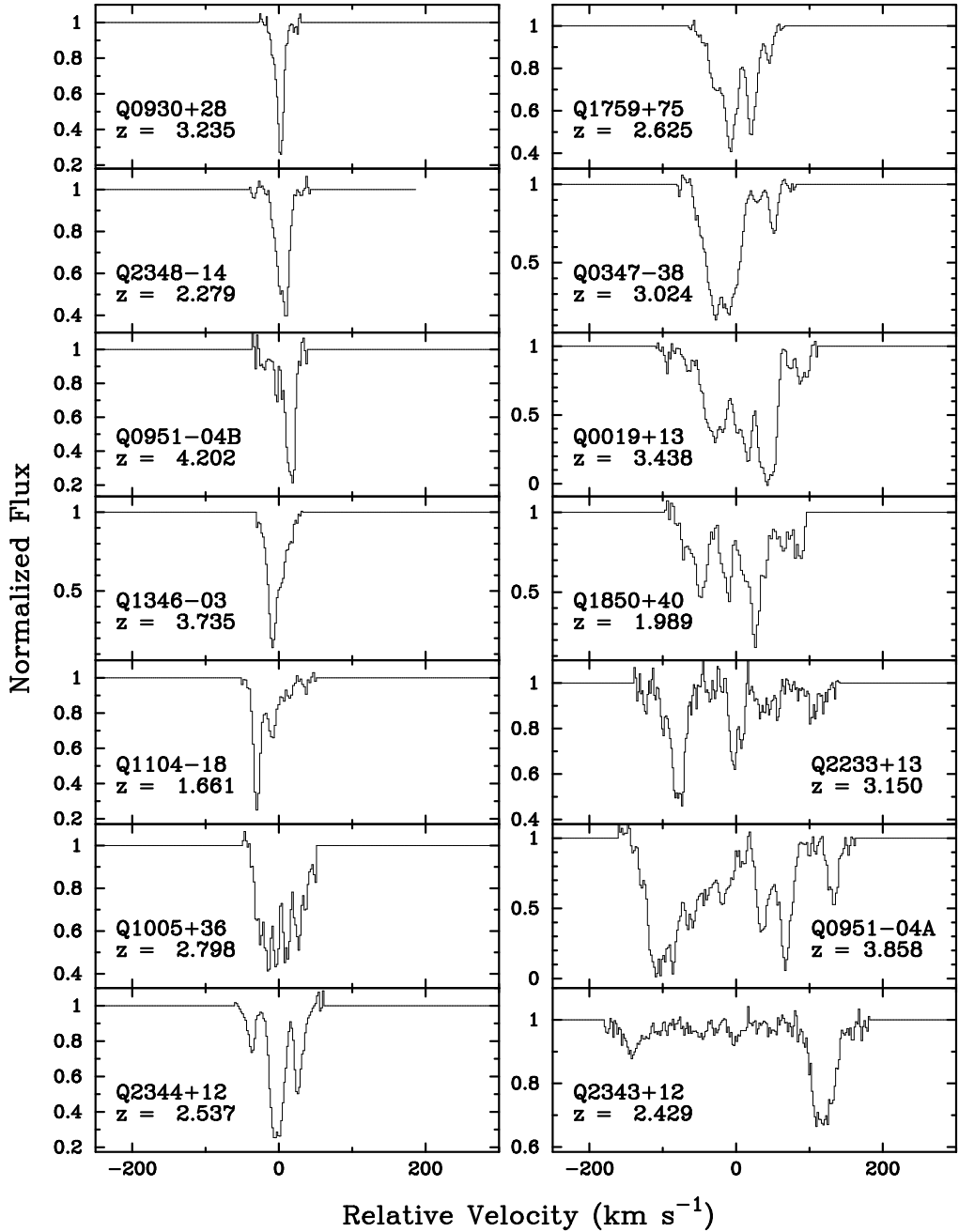}
\caption{Velocity profiles of low-ion transitions for the
14 new damped \lya systems. 
For each profile, $v = 0$ \kms corresponds to the redshift
labeled in the plot. \label{newdat}}
\end{center}
\end{figure*}

\begin{figure*}
\begin{center}
\includegraphics[height=8.0in, width=6.0in]{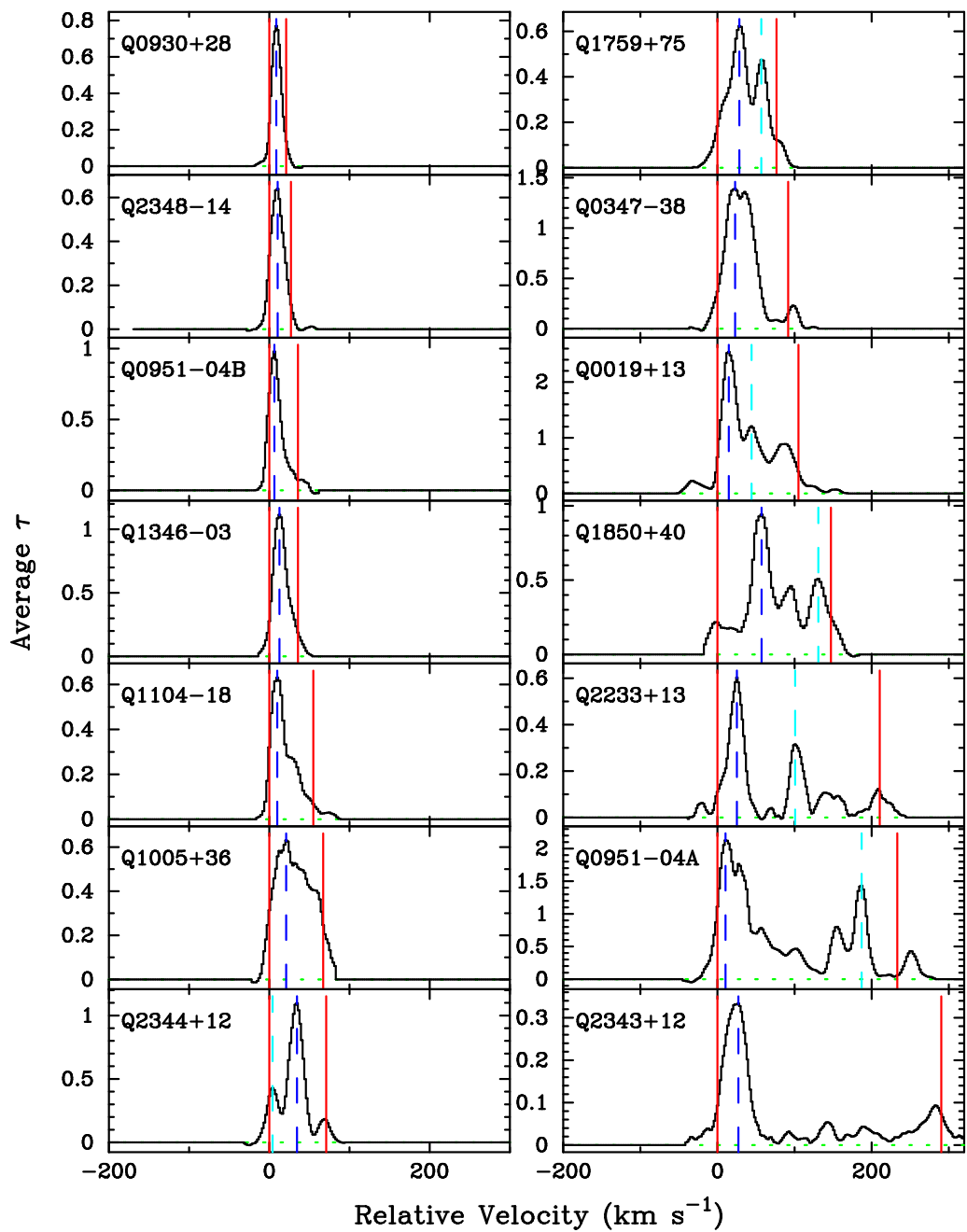}
\caption{Average optical depth profiles (binned over 9 pixels)
for the 31 transitions comprising our new total empirical data set.
The solid vertical lines designate the velocity interval, $\delv$,
while the dash-dot lines indicate peaks and the dotted line is the
$2 \sigma(\bar\tau)$ array.  
The profiles have been shifted in velocity space such
that the left edge of the velocity interval coincides with $v=0$ and
reflected in about half of the cases
so that the strongest peak always lies on the left edge of the interval.
\label{newtau}}
\end{center}
\end{figure*}

\clearpage

\noindent
several of these damped systems no profile 
satisfies all the selection criteria of PW.
Because we wish to 
maximize $\bar \tau(v) / \sigma(\bar\tau)$ 
at the edge of each profile, in particular to 
ensure an accurate measurement
of the velocity width,
we adopt several profiles with peak normalized flux 
$\Ipk < 0.1$, which violates the previously
established selection criterion (i.e.\ as in PW).  
In addition, we have relaxed the criterion that $\btau \geq 20$
on the grounds that numerical tests indicate this criterion
was too conservative.
In Figure~\ref{newdat} 
we present the velocity profiles as normalized intensity
versus velocity, and the corresponding binned optical
depth arrays in Figure~\ref{newtau}.
Note that many of the new profiles (e.g. Q0019$-$01, Q1104$-$18)
exhibit the same 
edge-leading asymmetry characteristic observed for the profiles of
Sample A.  
Furthermore, the distribution of profile widths 
resembles that of Sample A, but the new distribution extends to 290 \kms
which exceeds the 200 \kms maximum width in Sample A.
In fact, three of the 14 profiles have $\delv > 200 \mkms$.
The incidence of $\delv > 200 \mkms$ in only one of 17 previous
velocity profiles is due to small number statistics.

\begin{figure}[ht]
\begin{center}
\includegraphics[height=3.5in, width=3.0in, angle=-90]{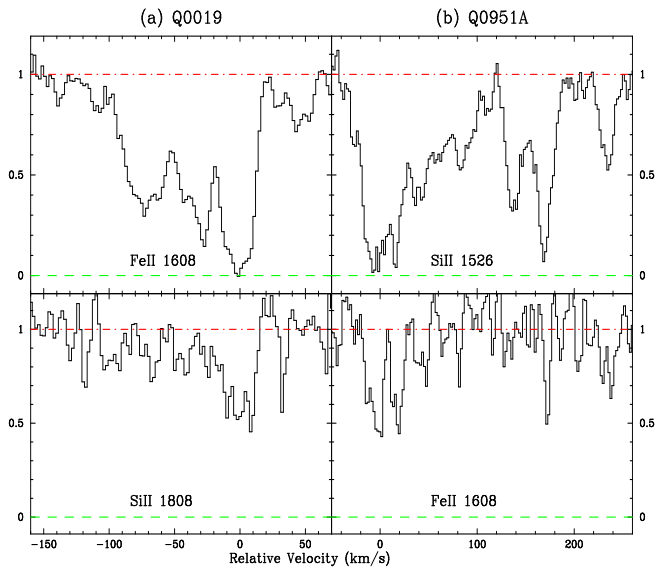}
\caption{Comparison of a mildly saturated low-ion
profile with a weak profile for the damped \lya systems toward 
(a) Q0019$-$01 at $z=3.439$ and (b) Q0951$-$04 at $z=3.859$. 
Note the profiles track one another very closely and therefore exhibit
nearly the same kinematic characteristics, irrespective of the
saturation.
\label{satfig}}
\end{center}
\end{figure}

Sample B includes two profiles with $I(v)/I_c < 0.1$,
including one with $I(v)/I_c < 0.0$ where the flux falls below zero
due to the effects of sky subtraction and Poisson noise.
As emphasized in PW, the statistical tests 
used to characterize the profiles are sensitive to
saturation;  for example, a false determination of $v_{pk}$
leads to incorrect values for all 4
statistical tests.
We have strong reason to believe, however,
that we are not contaminating the results by including these two profiles.  
First, in each case the profile has $I(v)/I_c < 0.1$ 
for only a few pixels.  Therefore, the centroid of the 
peak is accurately determined
to within a few \kms.  Secondly, we can compare the saturated
profiles with weaker, noisier, profiles from the same damped system.
In Figure~\ref{satfig}, we show this comparison.  Note that the profiles
track one another very closely
exhibiting nearly identical kinematic characteristics.
Therefore, the kinematic results are
not compromised by the minimal saturation evidenced in these two 
profiles.  To calculate $\bar\tau (v)$ 
over the interval where $I(v)/I_c < 0.0$
we arbitrarily 
set $I(v)/I_c = \sigma(I)/2$ and calculate $\bar\tau(v)$ accordingly.
Having performed numerical experiments which demonstrate that the
original criteria were too strict,
we establish a new set of saturation criteria:
(1) the line profile must be free of blending with other absorption
line profiles,
(2) the profile must not saturate for an excess of 1 resolution element, 
(3) the saturated region must not exceed one-fourth
of the profile velocity width, 
(4) $\btau \geq 10$ instead of the previous value of 20.
Together these criteria prevent
an inaccurate determination of $v_{pk}$ and any resulting
error in the test statistics.

To facilitate any parallel analysis performed
with the damped \lya surveys, where an
accurate $\N{HI}$ is always measured (e.g. \cite{wol95a}),
and to allow for the most accurate comparison with the Monte Carlo simulations,
we construct a subset of Sample B by imposing a stricter
$\N{HI}$ criterion than that of PW.
We refer to this subset of Sample B as Sample C.
The criterion requires 
an $\N{HI}$ measurement for every damped system which must exceed
$N_{thresh} = 2\sci{20} \cm{-2}$.  This eliminates two 
systems from Sample A (Q1946$+$60A and Q0449$-$13) 
inferred to be damped \lya systems on account
of their large ionic column densities and one system (Q2212$-$16) with a 
measured $\N{HI} < N_{thresh}$.  In addition, the criterion eliminates
Q2233$+$13 from the new systems.  
Therefore, Sample C contains profiles from 27 damped systems.
To investigate the evolution of the kinematic characteristics of the
damped \lya systems with redshift, we divide
Sample C at its median redshift and consider two subsets:
(i) a low redshift set of profiles with $\bar z = 2.06$
and $z_{median} = 1.96$ referred to as Sample D and
(ii) a high redshift set of profiles with $\bar z = 3.24$
and $z_{median} = 3.35$ referred to as Sample E.  Table~\ref{tab_sample}
summarizes the 5 data samples.

\begin{table}[ht] \footnotesize
\caption{\label{tab_sample}}
\begin{center}
{\sc Data Samples \smallskip} 
\begin{tabular}{ccl}
\tableline
\tableline \tskip
Sample & $N_{sys}$ & Comment \\
\tableline \tskip
A & 17 & Original selection criteria \nl
B & 31 & New selection criteria \nl
C & 27 & $\N{HI}$ measurement required \nl
D & 14 & Low Redshift Cut of Sample C\nl
E & 13 & High Redshift Cut of Sample C\nl
\tskip \tableline
\end{tabular}
\end{center}
\end{table}

\begin{figure}[hb]
\begin{center}
\includegraphics[height=5.0in, width=3.5in, bb = 65 58 547 734]{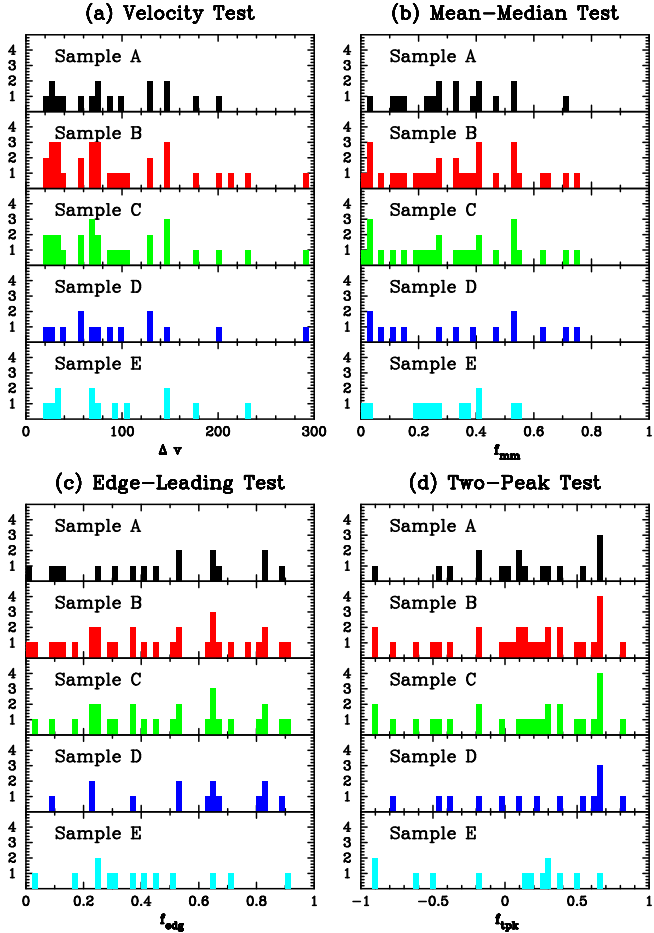}
\caption{Statistical test distributions for data Sample 
(A) the 17 damped \lya profiles from PW,
(B) the 31 profiles comprising the current data sample,
(C) 27 of the current profiles which satisfy stricter $\N{HI}$ criteria,
(D) a low redshift sub-sample and (E) a high redshift sub-sample.
\label{fig_smpl}}
\end{center}
\end{figure}

Figure~\ref{fig_smpl} shows the statistical test distributions for 
Samples A$-$E.  
The results for Samples A and B are in good agreement indicating that the
addition of the new systems will serve to strengthen the conclusions of PW.
The only significant difference between the two samples is the tail observed
out to nearly 300 \kms in the $\f{\delv}$ distribution of Sample B.
One also notes Samples B and C exhibit nearly identical distributions, 
therefore the stricter $\N{HI}$ criterion has little
effect on our results.
Similarly, the only disagreement observed by eye between
Samples D and E is in the Mean-Median Test, but 
statistically the two distributions are consistent ($P_{KS} = 0.6$).
We argue, therefore, that there is little evidence for an evolution of
the kinematic characteristics with redshift.  This will have significant
impact on interpreting the damped \lya systems in terms of the competing
galaxy formation scenarios and will be
further addressed in a future paper.

\subsection{ORIGINAL MODELS REVISITED}
\label{revisit}

\begin{figure}[ht]
\includegraphics[height=5.0in, width=3.5in, bb = 65 58 547 734]{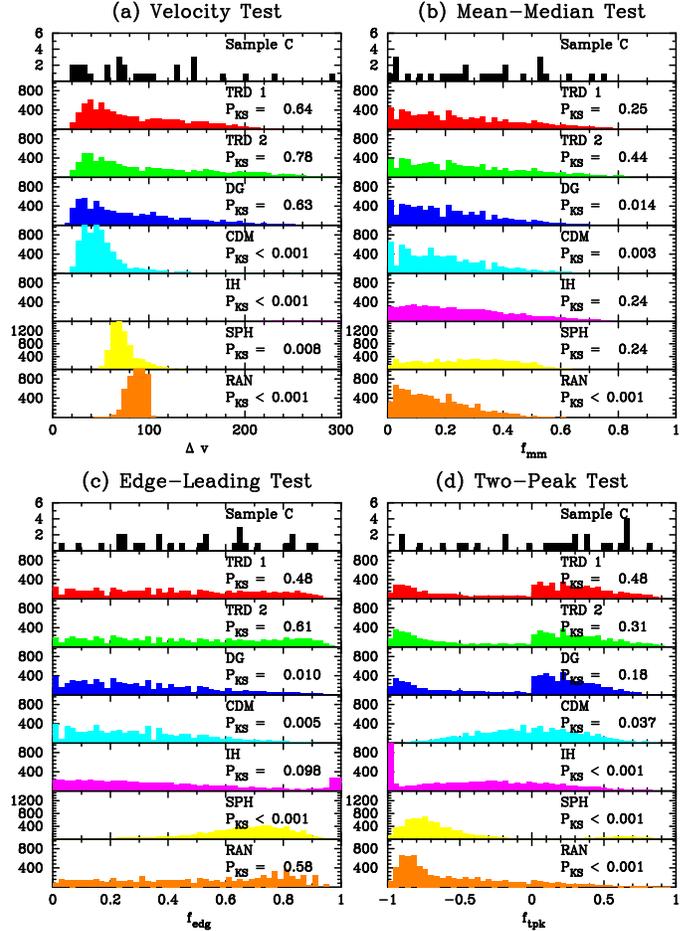}
\caption{The statistical test distributions for the models introduced
in PW compared against Sample C.  Note that all of the models
which were ruled out in PW ($P_{KS} < 5\%$) are ruled out at even higher
confidence levels now.
\label{fig_mdls}}
\end{figure}

Figure~\ref{fig_mdls} presents the statistical test distributions
for Sample C as well as the distributions for 7 of the Monte Carlo
models presented in PW.   The $P_{KS}$ values are the KS Test probabilities
that the model distribution and the corresponding distribution
from Sample C could have been drawn from the same parent population.
The TRD1 model is the 'best fit' model to Sample A.  For this model
to $h/R_d = 0.3$, $\Nperp = 10^{21.2} \cm{-2}$ and
$v_{rot} = 225 \mkms$,   where $h$ is the vertical scale length,
$R_d$ is the radial scale length, $\Nperp$ is the
central column density normal to the disk, and $v_{rot}$ is
the flat rotation speed.  The TRD2 model is identical to the TRD1 model
except $v_{rot} = 300 \mkms$.  
The remaining models are identical to those from $\S 6$ of PW.

\begin{figure}[ht]
\includegraphics[height=5.0in, width=3.5in, bb = 45 38 567 754]{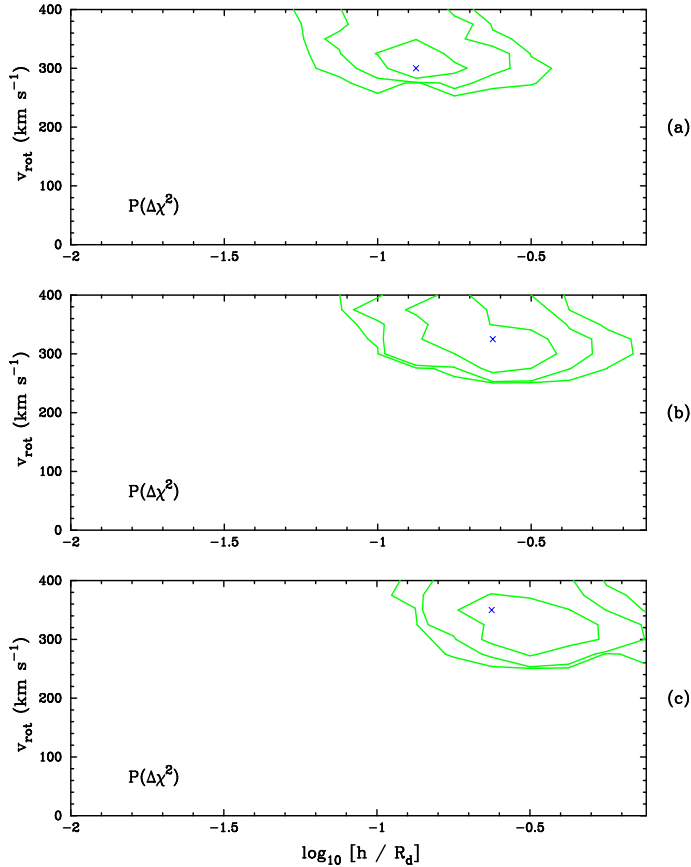}
\caption{Isoprobability contours derived from the Likelihood
Ratio Test for the TRD model against the $f_\delv$
distribution from Sample C with
$\Nperp = $ (a) $10^{20.8} \cm2$, (b) $10^{21.2} \cm2$, and 
(c) $10^{21.6} \cm2$.
Contour levels are drawn at $P$  = 0.01, 0.05 and 0.32.
Note that the $f_\delv$ distribution would imply rotation
speeds $v > 250 \mkms$ given the assumption of a single
population of disks. \label{mctrd}}
\end{figure}

The addition of the new profiles has tightened the primary results of PW
in every case. In particular, we 
point out the continued failure of every model except the TRD
model to reproduce the damped \lya observations.
The TRD1 model is generally consistent with Sample C, but
because the model adopts the same rotation speed
($v_{rot} = 225 \mkms$) for all of the disks, it   
predicts no damped \lya systems with velocity width $\delv > 250 \mkms$.
Hence, this simple model cannot reproduce the
high $\delv$ tail of the
new empirical $\f{\delv}$ distribution.
By contrast, the distribution
from the TRD2 model 
extends to $\delv \approx 300 \mkms$ and therefore provides a better
match to the high $\delv$ tail.
Figure~\ref{mctrd} plots the relative likelihood ratio
test results for Sample C against the TRD model with the central
HI column density (a) $\N{HI} = 10^{20.8} \cm{-2}$, 
(b) $\N{HI} = 10^{21.2} \cm{-2}$, and (c) $\N{HI} = 10^{21.6} \cm{-2}$.
The figure indicates a lower limit of 
$v_{rot} > 250 \mkms$ and that the optimal
rotation speed is $v_{rot} \approx 300 \mkms$.
However, if damped systems evolve to present galaxies, then a single
population of disks with $v_{rot} = 300 \mkms$ 
is unacceptable as such large rotation
speeds are rarely observed. 
One concludes the single
population disk model is too simple and must be revised.
Models including distributions of rotation speeds and sightline
encounters with multiple disks will be considered in future
papers.  In this paper we retain the assumption of a single population
of disks and study how more realistic disk properties affect
the test statistic distributions.  

\section{IMPROVED DISK MODELS}

In this section we improve on the TRD model by introducing more
realistic disk characteristics.  We test the TRD model against the
damped \lya observations in the light of these properties and thereby gauge
the robustness of the model.
The discussion focuses primarily on the velocity width
test statistic because the other test statistics are less
sensitive to changes in disk properties.

\subsection{THE WARPED DISK MODEL}

HI 21 cm observations of local spiral galaxies
(e.g. \cite{brg90}) suggest
a significant fraction of disks are warped.
Therefore, if the damped \lya systems evolve
into disk galaxies, as we have argued,
the effects of warps must be considered.
Warps produce two opposite kinematic effects: (i)
some warped disks will have a large cross-section to sightlines
which satisfy $\N{HI} \geq 2 \sci{20} \cm{-2}$ with large impact
parameter and therefore small $\delv$.  For instance,  a tangent
to the outer edge of a warped disk may have very large impact
parameter (small $\delv$) yet still
satisfy the $\N{HI}$ criterion (see Figure~\ref{warpex}a below), 
and (ii) those sightlines which doubly penetrate
the disk (at a warped edge in addition to the unwarped inner disk,
see Figure~\ref{warpex}b)
will tend to yield larger $\delv$ than
if they penetrated an unwarped disk.  If the latter effect
dominates, one expects 
a warped disk to yield a $\f{\delv}$ 
distribution with a greater number of large $\delv$ values,
which imitates a thicker and/or more rapidly rotating
unwarped disk.  In some cases, however, we find that the
former effect dominates the results thereby lowering the
median of the $\f{\delv}$ distribution.

\begin{figure*}
\begin{center}
\includegraphics[height=6.0in, width=3.7in, angle=-90,bb = 55 48 557 744]{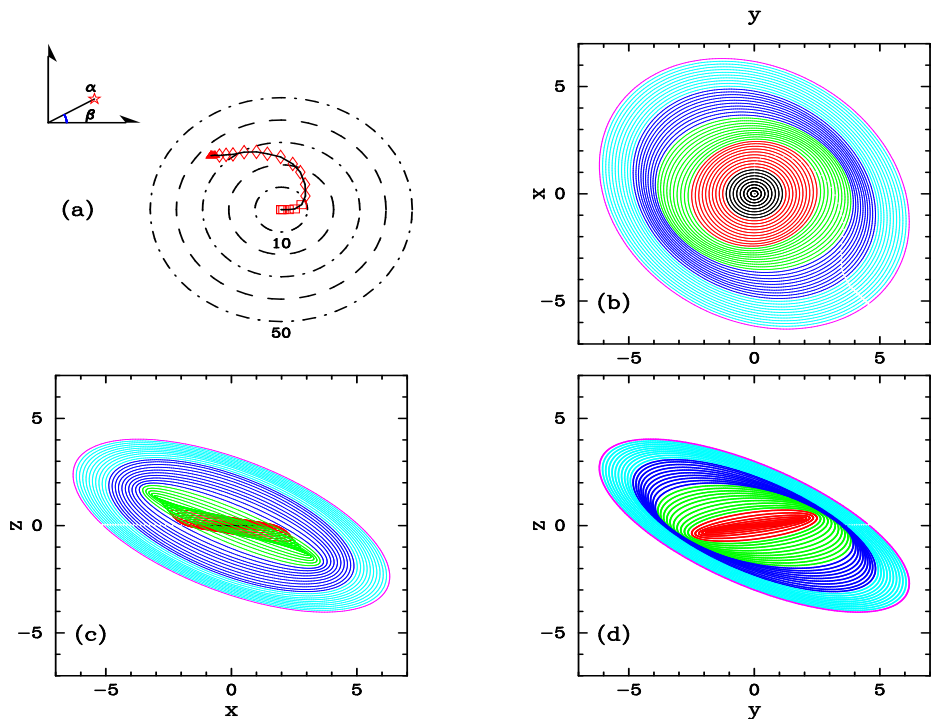}
\caption{The polar plot depicts the line of nodes for a warped disk.
Each successive point on the solid line indicates rings of 
monotonically increasing radii.  The azimuthal angle ($\beta$) indicates
the rotation angle of the ring 
and the `radial' angle ($\alpha$) designates the tilt angle of
the ring.  The remaining plots show the warped disk as a series of rings
viewed from three perspectives.
\label{lonex}}
\end{center}
\end{figure*}

\begin{figure*}
\begin{center}
\includegraphics[height=6.0in, width=3.7in, angle=-90,bb = 55 48 557 744]{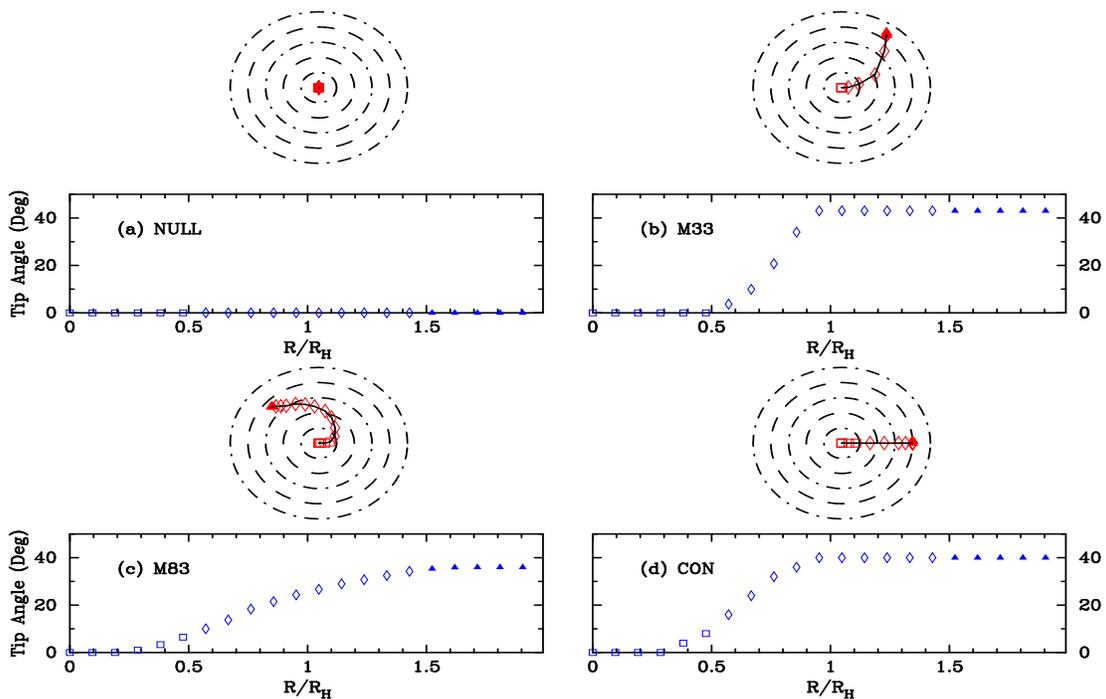}
\caption{Lines of nodes for the four canonical warped disks:
(a) an unwarped disk, (b) M33, (c) M83, and (d) an artificial warp.
The polar plot has the same meaning as in Figure 7.  The x-y
plot reveals the radial dependence of the tilt angle as a function
of the effective Holmberg Radius, $R_H$.
\label{lons}}
\end{center}
\end{figure*}

\begin{figure*}
\begin{center}
\includegraphics[height=8.0in, width=6.0in]{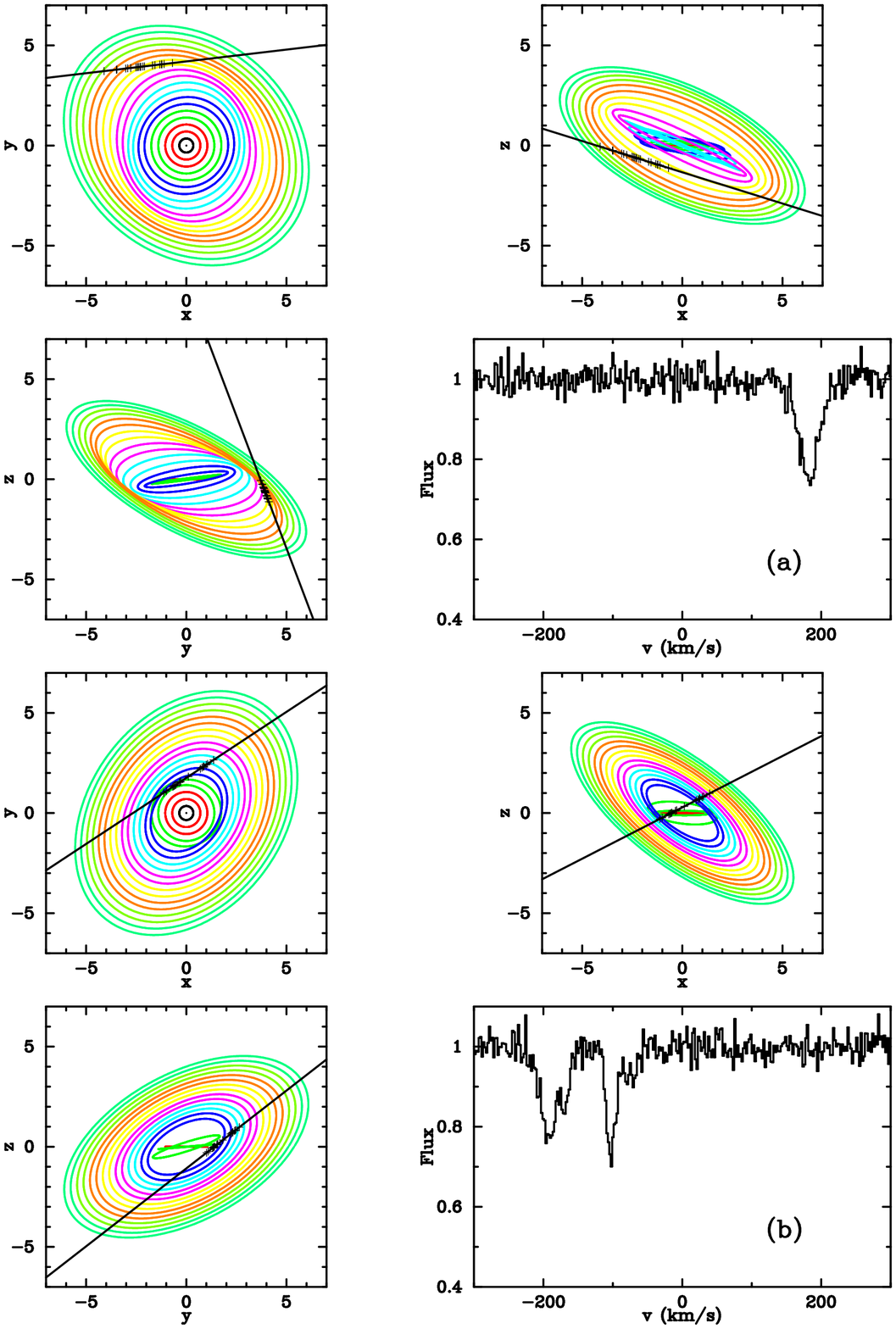}
\caption{Illustration of the two principal kinematic effects associated
with warped disks:  (a)  a sightline with large impact parameter
yielding a small $\delv$ which would not be included in a treatment
of an unwarped disk as $\N{HI}$ would be less than $N_{thresh}$;
(b) a sightline doubly penetrating the disk resulting in a significantly
larger $\delv$ than if the disk were not warped.  The + signs indicate
the position of clouds along the sightline.
\label{warpex}}
\end{center}
\end{figure*}

Unlike the TRD model,  the warped disk model
cannot be treated analytically.  Standard modeling techniques involve
decomposing the disk into a number of concentric rings, 
each with a unique
orientation, with the configuration described by
a 'line of nodes'.  Figure~\ref{lonex}a is a polar plot of a 'line
of nodes' where each successive point identifies 
rings with larger radii.  We have used
different symbols to help distinguish
the points.  The radial distance ($\alpha$)
to each symbol indicates the tilt angle, and
the azimuthal position ($\beta$)
designates the position angle of the tilt axis.
We define the reference plane by the
orientation of the rings with smallest radius, i.e.\ the inner disk.
Figures~\ref{lonex}b-d show 3 views of the
warped disk corresponding to this line of nodes.  
In the following, we consider 4 lines of nodes characterized by
$R_H$, an effective Holmberg Radius, which sets
the radial separation of the rings.
Figure~\ref{lons} plots the lines of nodes with respect to $R_H$.
Included is a NULL line of nodes (Figure~\ref{lons}a) 
describing an unwarped disk,  the lines of 
nodes for M83 and M33 (Figure~\ref{lons}b,c respectively; \cite{brg90}) and 
an artificially constructed line of nodes (CLON).
Because we treat the warp model numerically,
we will compare against
the results of the null models and not the analytic TRD Model
in order to properly measure the consequences of warping without
being biased by any numerical effects.
Figure~\ref{warpex} illustrates the two kinematic effects mentioned
above.  In Figure~\ref{warpex}a a sightline penetrates a disk described
by the WRP4 model and has a large impact parameter.  The column density
derived for this sightline is $\N{HI} = 1.6 \sci{21} \cm{-2}$ and
the resulting profile is plotted in the lower right panel.  Note
that this system would have $\N{HI} < 2 \sci{20} \cm{-2}$ for all 
of the other lines of nodes  considered here and would therefore
not contribute to the test statistic distributions.
The more common kinematic effect is illustrated in Figure~\ref{warpex}b
where we plot a sightline doubly penetrating a disk from the
WRP2 model.  The resulting profile is significantly wider than
if the disk were not warped.

\begin{table}[ht] \footnotesize
\caption{\label{wrpprm}}
\begin{center}
{\sc Warp Parameters \smallskip}

\begin{tabular}{lccc}
\tableline
\tableline \tskip
Label & $h/R_d$ & LON & $R_H/R_d$ \\
\tableline \tskip
WRP1 & 0.2 & NULL & --  \nl
WRP2 & 0.2 & M33  & 2.0 \nl
WRP3 & 0.2 & CLON  & 2.0 \nl
WRP4 & 0.2 & M83  & 2.0 \nl
WRP5 & 0.2 & CLON  & 3.0 \nl
WRP6 & 0.3 & NULL & --  \nl
WRP7 & 0.3 & M83  & 2.0 \nl
WRP8 & 0.3 & CLON  & 2.0 \nl
\tskip \tableline
\end{tabular}
\end{center}
\end{table}

\begin{figure}[hb]
\begin{center}
\includegraphics[height=5.0in, width=3.5in, bb = 55 48 557 744]{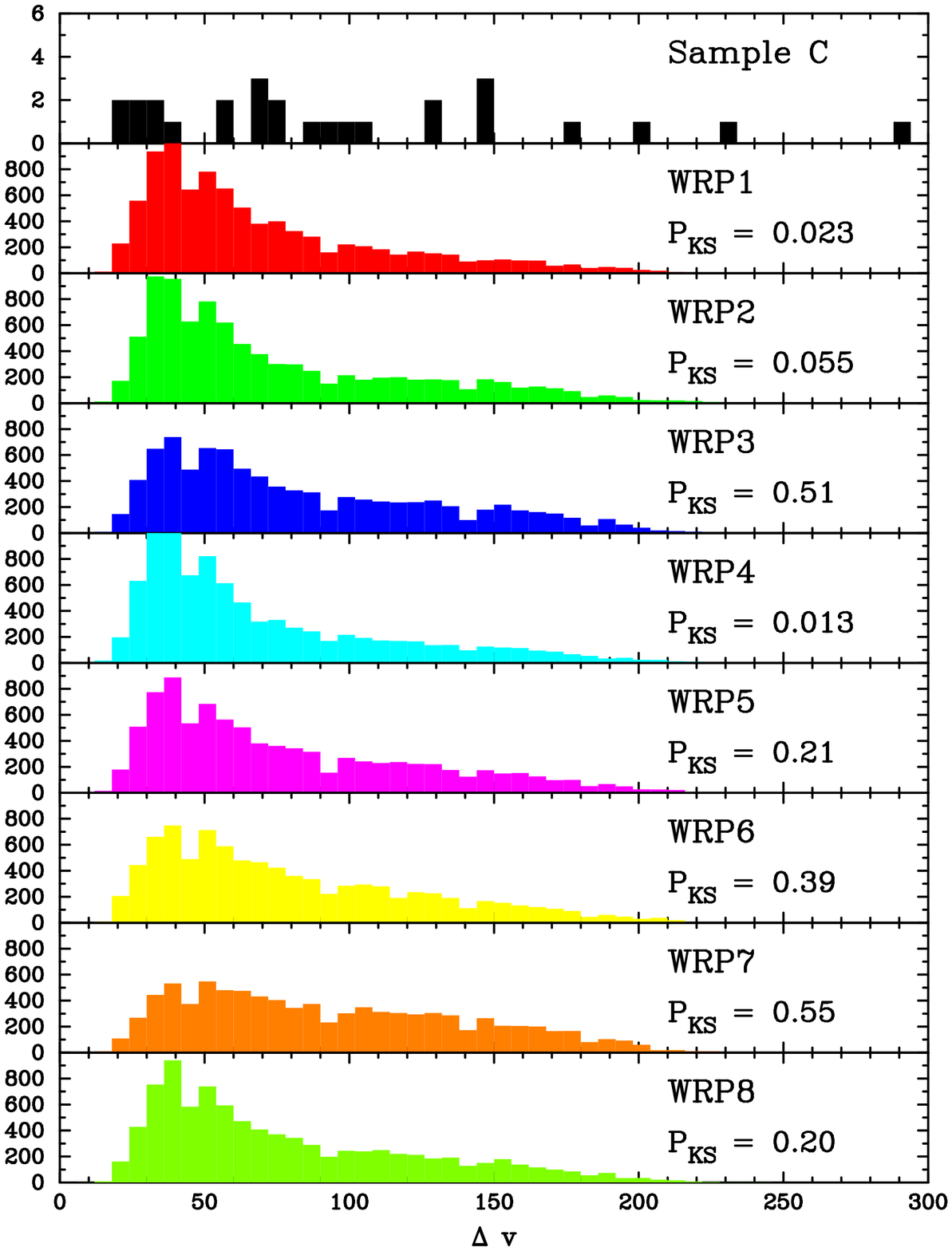}
\caption{The $\f{\delv}$ distribution for 8 warp models compared
against that for Sample C.  The results demonstrate that warping
has only a minor effect on the TRD Model.
\label{wrprst}}
\end{center}
\end{figure}

We performed a series
of Monte Carlo simulations for warped disks with a variety of thicknesses,
$R_H/R_d$ values, and lines of nodes.
Figure~\ref{wrprst} plots the $\f{\delv}$
distribution for 8 representative warp models corresponding to
the parameters listed in Table~\ref{wrpprm}. 
All of the models assume a rotation speed of $v_{rot} = 250 \mkms$
and a central column density, $\Nperp = 10^{21.2} \cm{-2}$. 
In general, warping yields a greater number of moderate
$\delv$ values ($\delv \approx \ohf v_{rot}$) and a few more 
large $\delv$ values. 
This is the case for the M33 and CLON lines of nodes (e.g.\ compare the null
model, WRP1, with
WRP2 and WRP3) where the results resemble those for a thicker disk.
This is not true for all lines of nodes.
Note the difference between the results for models WRP3
and WRP4 which correspond to the CLON and M83 lines of nodes
respectively.  While both warps attain a tilt angle of $\approx 40^\circ$
at $R_H = 2 R_d$, the M83 warp has a much larger cross-section to
sightlines satisfying the $\N{HI}$ criterion.  As demonstrated in PW, those
sightlines with large impact parameters tend to have smaller $\delv$,
therefore the WRP4 model actually does more poorly than the unwarped
disk, even though there are a number of sightlines which
doubly penetrate the disk.
The results are also sensitive to the adopted value of $R_H/R_d$.
Comparing models WRP3 and WRP5 one notes the reduced effects of warping 
for larger $R_H/R_d$.  As typical values of $R_H/R_d$
in local disks generally exceed 3, we expect warping to have even
less of an effect on the $\f{\delv}$ distribution than presented here.
Curves WRP6$-$8 demonstrate the results for warping for thicker
disks.  Qualitatively, the changes in the $\f{\delv}$ distribution
are the same as those for the thinner disks.

To summarize, warping has only a moderate effect on the
profile kinematics.  For those lines of nodes
where the dominant effect is of sightlines that penetrate the disk
two times,
warping mimics thicker disks, i.e.\ one observes a larger number
of moderate $\delv$ values in the $\f{\delv}$ distribution.
Furthermore, warping can also lead to significantly fewer moderate
$\delv$ values when the cross-section to damped \lya system
extends to higher impact parameter.  This is the case for systems where
the outer rings of the warp have $\beta \approx 180^\circ$.
Having performed simulations for a large range of $h$, $R_H/R_d$,
and lines of nodes, we quantify the results as follows:
(1) in extreme cases, warping mimics disks with up to
$50\%$ larger or smaller effective thickness ($h/R_d$ value),
(2) warping lends to few additional large $\delv$ values in the 
$\f{\delv}$ distribution and
therefore has little effect on the acceptable values for $v_{rot}$, 
and (3) we find very thin ($h/R_d < 0.1$) warped disks are inconsistent
with the damped \lya observations.

\subsection{ROTATION CURVES}

In PW, the velocity field of
the exponential disk model was given by $v_\phi = v_{rot}, v_R=v_Z = 0$,
i.e., we assumed a flat rotation curve independent 
of radius and height above the midplane of the disk.  
This is a good zeroth order representation for the rotation curve
of spiral disks at large radii.  
But it breaks down at small radii because of the singularity implied
for the mass density at $R= 0$.  In more realistic models the density
approaches a finite central value at $R$ less then some core 
radius, resulting in $v_\phi = 0$ at $R=0$.  The independence of 
$v_\phi$ with $Z$ may also be unrealistic.  In their analysis of
the kinematics of the ionized gas associated with 
CIV absorption lines, Savage and Sembach (1995) find evidence
that $v_\phi = v_{rot}$ up to $|Z| \approx 5$ kpc above 
midplane.  On the other
hand Sancisi (1998) finds evidence for a decrease in $v_\phi$
with increasing $|Z|$ in his
HI studies of edge-on spirals.  In fact, unless there
is strong coupling between layers of adjacent $Z$ this is what one 
expects. Therefore, a more physical rotation curve will (i) approach
$\vrot = 0 \mkms$ at $R=0$ where the enclosed mass density presumably
approaches a finite value and (ii) likely decrease in speed with
increasing height above the midplane. In this section, we 
consider a range of rotation curves
appropriate for a system comprised of a thick disk, bulge, and dark matter
halo.  

\subsubsection{Thick Disk}

As demonstrated in PW (reaffirmed in $\S$\ref{revisit}),
the kinematics of the damped \lya systems
are consistent with thick rotating disks. 
Furthermore, we showed that the vertical scale height, $h$,
must be greater than one-tenth the radial scale length, $R_d$.
Therefore, the thin disk approximation is not applicable
for deriving the rotation curve of these disks.
In this section, we detail a
numerical solution for the rotation curve of a thick disk by 
explicitly solving Poisson's equation and combining the solution
with the condition for centrifugal equilibrium.

\paragraph{The Fourier-Bessel Transformation Method}

Our approach is an extension of the technique developed by Toomre (1963)
for the thin disk solution and more recently applied by 
Casertano (1983) to the rotation
curve for a thick disk at midplane.  We 
calculate the disk potential by Fourier-Bessel transforming Poisson's
equation,

\begin{equation}
\nabla^2 \Phi = 4 \pi G \rho \perd
\label{poisson}
\end{equation}

\noindent We adopt an exponential mass density of the form,

\begin{equation}
\rho (R,Z) = \rho_0 \exp (-R / R_d) \exp (-|Z| / h) \cmma
\end{equation}

\noindent where $\rho_0$, the central mass density, is related
to the central surface density, $\Sigma_0$, and 
central HI column density, $\Nperp$,  by

\begin{equation}
\rho_0 = {\Sigma_0 \over 2 h} = \mu m_p {\Nperp \over 2 h} \cmma
\end{equation}

\noindent with $\mu$ the molecular weight relative to Hydrogen.
This relation implicitly assumes that the mass of the disk
is dominated by HI gas, i.e., we ignore any contribution from
stars or molecular Hydrogen. 

Taking the Fourier-Bessel transformation of Poisson's equation, 
we find

\begin{equation}
\tilde \Phi (k, Z) = {- 2 \pi G \over k}
\intl_{- \infty}^{+ \infty} \exp (- k |Z - \zeta|)
\, \tilde \rho (k,\zeta) \, d \zeta \cmma
\label{phitild}
\end{equation}

\noindent where


\begin{equation}
\tilde \rho(k, \zeta) = \rho_0 \; \exp \ltp {- |\zeta| \over h} \rtp
\times
\label{rhotild}
\end{equation}
\begin{displaymath}
\intl_0^\infty J_0 (kR) \; \exp (-R/R_d) \; R \; dR \perd
\end{displaymath}

\noindent  This integral is analytic (\cite{grd80}, 6.623 [2]):

\begin{equation}
\tilde \rho(k,\zeta) = {2 \, R_d^{-1} \, \Gamma(\rhf) \rho_0 \over 
\sqrt{\pi} \, (R_d^{-2} + k^2)^\rhf} \; \exp \ltp {- |\zeta| 
\over h} \rtp \perd
\label{rhotildB}
\end{equation}

\noindent Evaluating $\tilde \Phi (k,Z)$, we find 

\begin{equation}
\tilde \Phi (k,Z) = {A \over k \, (R_d^{-2} + k^2)^\rhf}
\bigg\{ {1 \over hk+1} \big [ \exp (-Z/h) 
\label{phitildB}
\end{equation}
\begin{displaymath}
+ \exp (-k Z) \big ] \, +
{1 \over hk-1} \ltk \exp (-Z/h) - \exp (-k Z) \rtk \bigg\} \perd
\end{displaymath}

\noindent where the constant $A$ is

\begin{equation}
A = {G \, \pi \, \Sigma_0 \over R_d} \perd
\end{equation}

\noindent Finally, we take the inverse transform of $\tilde \Phi$,

\begin{equation}
\Phi (R,Z) = \intl_0^\infty J_0 (kR) \; \tilde \Phi(k,Z) \; k \; dk \perd
\label{phiRZ}
\end{equation}

\noindent  The thick disk rotation curve is therefore given by,


\begin{equation}
v_\phi^2 |_{\rm disk} (R,Z) = - R  {\partial \over \partial R} \,
\Phi (R,Z) 
\label{vcRZ}
\end{equation}
\begin{displaymath}
= R \intl_0^\infty J_1 (kR) 
\, \tilde \Phi (k,Z) \, k^2 \, dk \cmma
\end{displaymath}

\noindent which can be evaluated very accurately with standard 
numerical techniques.  
As the value for $\Sigma_0$
derived from the damped \lya observations (i.e.\ $\Nperp$)
is significantly lower than that observed in present galaxies,
the disk rotation curve will peak at $v_{max} \ll 200 \mkms$
($v_\phi^2 \propto \Sigma_0 R_d$) unless
the HI gas contributes only a small fraction $(< 30\%)$ of the disk
mass.  While it is possible that stars may double the surface density
in the inner part of the disk (see \cite{wol98}), 
we proceed under the assumption that the HI gas dominates.

\begin{figure*}
\begin{center}
\includegraphics[height=6.0in, width=3.9in, angle=-90]{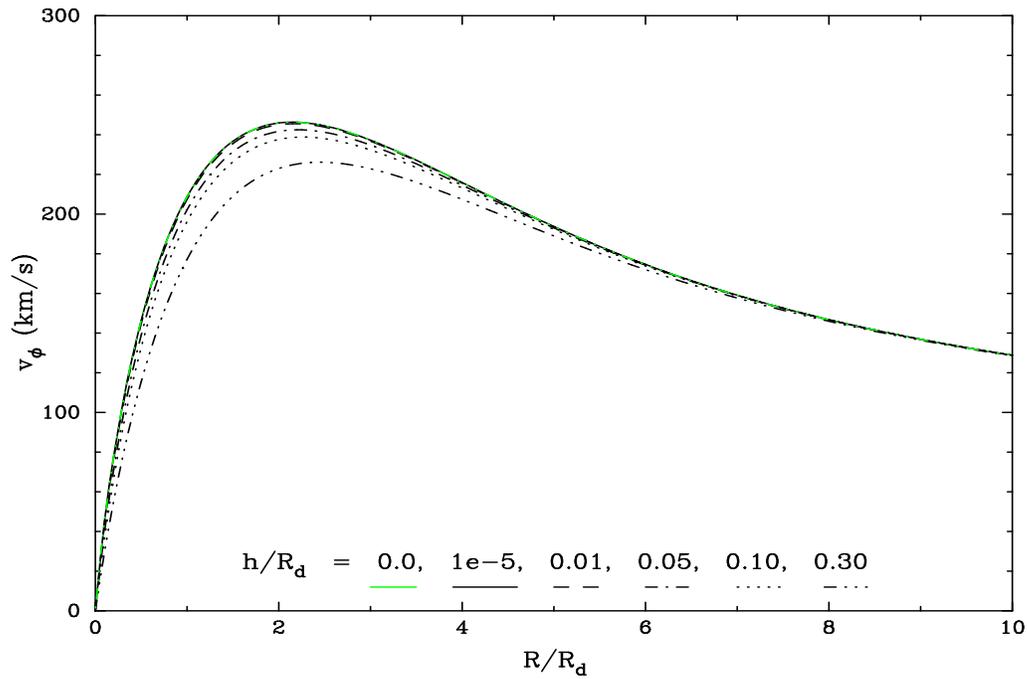}
\caption{The rotation curve at midplane, $v_{circ} (R,0)$, for five disks
with varying thickness, $h/R_d$.  All of the curves are normalized
to have the same central column density $\Nperp$.  There is only
small dependence on thickness for the rotation curves.
\label{hcomp}}
\end{center}
\end{figure*}

\begin{figure*}
\begin{center}
\includegraphics[height=6.0in, width=3.9in, angle=-90]{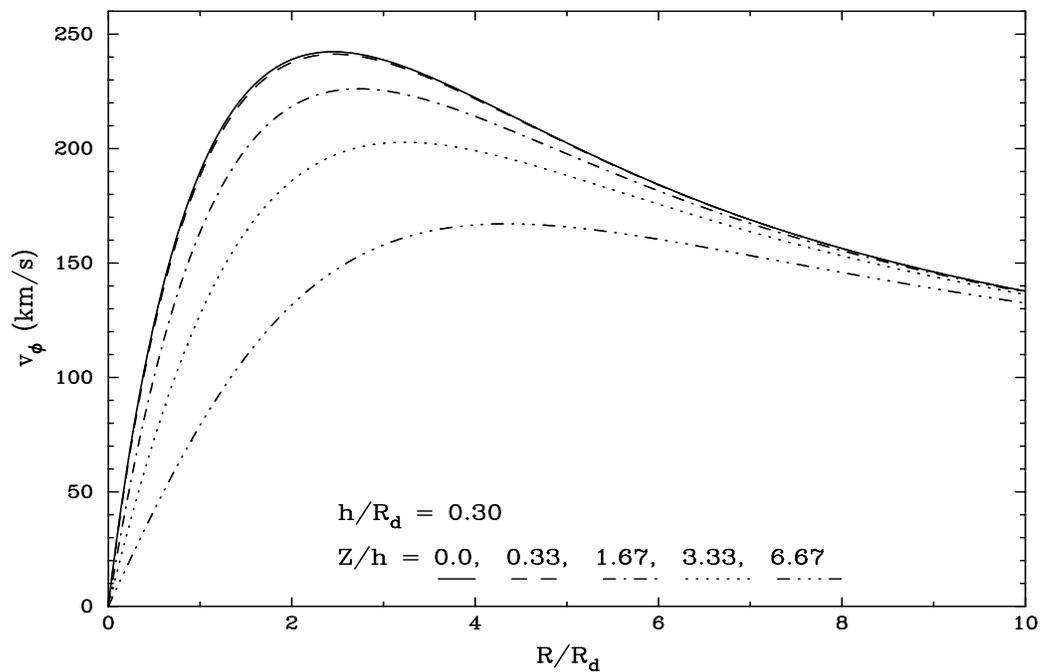}
\caption{The rotation curves for a disk with $h/R_d = 0.3$ at
various heights within the disk.
\label{zcomp}}
\end{center}
\end{figure*}

\begin{table*}
\caption{\centerline{\sc Rotation Curves \smallskip} 
\label{tab_rot}}
\begin{center}
\begin{tabular}{cccccccl}
\tableline
\tableline \tskip
Crv & 
$v_{disk}^{max}$ & $\Sigma$ &
$v_b$ & $R_b$ & 
$v_h$ & $R_h$ & Comment \\
& (km/s) &  $(\msol / {\rm pc^2})$
& (km/s) & ($R_d)$
& (km/s) & ($R_d)$ \\
\tableline \tskip
1 & 250 & -- & --  & --  & --  & --  &  FLT \nl
2 &  41 & 18 & -- & -- & -- & -- & DLA disk \nl
3 & 250 & 1450 & -- & -- & -- & -- & Massive disk \nl
4 & 250 & 18 & -- & -- & 250 & 0.5 & DSK+HLO 1 \nl
5 & 250 & 18 & -- & -- & 250 & 5.0 & DSK+HLO 2 \nl
6 & 250 & 18 & 180 & 0.1 & 250 & 0.5 & DSK+HLO+BLG 1 \nl
7 & 250 & 18 & 180 & 0.1 & 250 & 5.0 & DSK+HLO+BLG 2 \nl
\tskip \tableline
\end{tabular}
\end{center}
\end{table*}

\paragraph{Thick Disk Curves}

Figure~\ref{hcomp} shows the rotation curve at midplane,
$v_\phi (R,0)$, derived from Equation~\ref{vcRZ}
for 5 disks with varying thickness.  Also plotted is the
rotation curve derived with the thin disk approximation, normalized
to have the same maximum rotation speed
as the curve with $h/R_d = 1\sci{-5}$. These two 
curves are in such close agreement that they are indistinguishable
in Figure~\ref{hcomp}, verifying the accuracy of the numerical solution.  
Note all of the other curves are normalized 
to have the same central column density, $\Nperp$.
Physically this normalization
implies very different central mass densities for the different
thickness disks ($\rho_0 \propto \Nperp h^{-1}$). 
There is a decrease in $v_\phi$
with increasing $h/R_d$, as expected, because $\rho_0$ is smaller
and therefore the centrifugal force is smaller at a given radius $R$.
Even for $h/R_d = 0.3$, however, the resulting rotation curve nearly
traces that for the thin disk.

The rotation curve at various heights in a disk with $h/R_d = 0.3$
is plotted in Figure~\ref{zcomp}.  Note 
the curves at $Z = 1.67 \, h$ and at midplane differ by less than $10\%$,
and even at $Z = 6.67 \, h$
there is generally less than a $50\%$ effect. 
We find similar results for disks of all thicknesses.
Since nearly all of the gas arises within $\approx 2h$ of the
midplane, in general we expect the decrease in $v_\phi$ with increasing
$Z$ to have little effect on the kinematic results.
In fact, it is possible the decrease in $v_\phi$ will actually
increase the differential rotation along a given sightline.

\subsubsection{Bulge and Halo}

For the bulge rotation curve we assume 
the Hernquist Bulge model (\cite{hern90})
parameterized by the peak velocity, $v_b$, and core radius,
$R_b$

\begin{equation}
v_{circ}^2 |_{\rm Bulge} (r) = 4 v_b^2 {r R_b \over (r + R_b)^2} \cmma
\end{equation}

\noindent
where we can relate $v_b$ and $R_b$ to the bulge mass, $M_b$  by the
following expression:

\begin{equation}
M_b = 2.7 \sci{9} \msol \; \ltp {v_b \over 200 \mkms} \rtp^2
\ltp {R_b \over 100 {\rm pc}} \rtp  \perd
\end{equation}

Meanwhile, for the halo we require a density profile which yields
a flat rotation curve at large radii.  
We adopt the following density profile,

\begin{equation}
\rho|_{\rm Halo} (r) = {\rho_h \over 1 + (r / R_h)^2} \cmma
\label{rhohalo}
\end{equation}

\noindent where $\rho_h$ is the central density of the halo
and $R_h$ is the core radius.  
This gives the following rotation curve,

\begin{equation}
v_{circ}^2 |_{\rm Halo} (r) = v_h^2 \ltk 1 - {R_h \over r} 
\tan^{-1} \ltp {r \over R_h} \rtp \rtk \cmma
\end{equation}

\noindent where $v_h$ is the halo rotation speed at large radii.

As both the bulge and halo are
spherically symmetric, the rotation curves in cylindrical 
coordinates are: 

\begin{equation}
v_\phi^2 |_{\rm Bulge} (R,Z) 
= {R^2 \over \sphr} {4 v_b^2 R_b \over (\sphr + R_b)^2} 
\label{vcbul}
\end{equation}

\noindent and 


\begin{equation}
v_\phi^2 |_{\rm Halo} (R,Z) = v_h^2 {R^2 \over R^2 + Z^2} 
\Bigg [ 1 -
\label{vchalo}
\end{equation}
\begin{displaymath}
{R_h \over \sphr} \tan^{-1} \ltp {\sphr \over R_h}
\rtp \Bigg ] \perd
\end{displaymath}

\begin{figure*}
\begin{center}
\includegraphics[height=8.5in, width=5.7in, bb = 55 48 557 744]{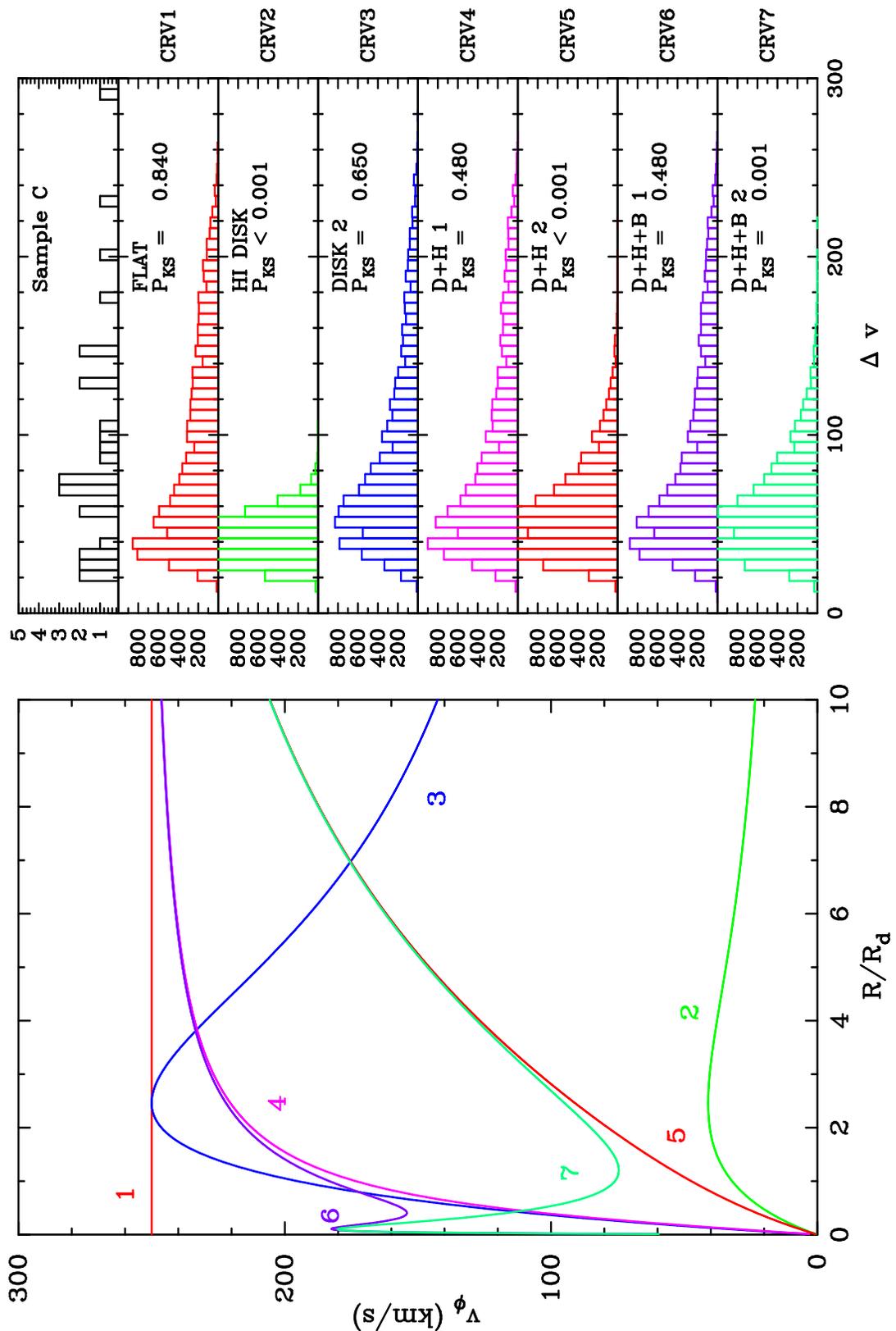}
\caption{The left hand side plots the rotation curve at midplane
for disks with varying bulge, halo, and disk components.  The right
hand side plots the $\f{\delv}$ distributions for the 8
curves compared against Sample C.
\label{rotrst}}
\end{center}
\end{figure*}

\subsubsection{Results}

The disk rotation curve is uniquely
parameterized by $h$ and $\Sigma_0$, where $h$ and all other distance
scales are in units of the radial scale length, $R_d$, and $\Sigma_0$ is
the disk surface density at $R=0$.
The bulge and halo
models are each parameterized by two variables which
describe the slope of the curve ($R_b, R_h$) and the maximum 
rotation speed ($v_b, v_h$) at $Z = 0$.  The total rotation speed, then, is
given by


\begin{equation}
v_\phi^2 |_{\rm Tot} (R,Z) = v_\phi^2 |_{\rm Halo} (R,Z) \; + \;
v_\phi^2 |_{\rm Bulge} (R,Z)
\end{equation}
\begin{displaymath}
 + v_\phi^2 |_{\rm Disk} (R,Z) \perd
\end{displaymath}

\noindent
We have performed simulations for
a wide range of the six parameters.  
The main results of the analysis are presented in
Figure~\ref{rotrst} and the rotational parameters
are given in Table~\ref{tab_rot}.  
Also listed is the peak rotation speed
of each rotation curve, $v_{max}$.
For all of the models presented here,
we assume $\Nperp = 10^{21.2} \cm{-2}$ and that $h/R_d = 0.3$.

To facilitate a direct comparison with the TRD model, we plot the
results for a flat rotation curve with $v_{max} = 250 \mkms$.
For a given value of $v_{max}$, 
the flat rotation curve yields a 
greater fraction of large velocity widths than any other rotation
curve.  Therefore, it gives
the best agreement with the empirical distribution provided
$v_{max} \lesssim 350 \mkms$, above which the distribution
has too many large $\delv$ values.
For model CRV2, we assume a thick disk with 
$\Sigma_0 = \mu m_p \Nperp = 18 \msol / {\rm pc^2}$
and $R_d = 10$ kpc which implies $v_{max} = 41 \mkms$. 
This model corresponds to $\Nperp = 10^{21} \cm{-2}$ as
inferred from the damped \lya observations. The value
of $R_d$ was chosen to roughly correspond to the cross-section
derived for damped \lya systems assuming a number density
approximately equal to that for spiral galaxies in the present
epoch (\cite{wol95a}).
As noted above and shown in Figure~\ref{rotrst}, 
the mass associated with the HI disk implies
a rotation curve which is
inconsistent with the damped \lya observations.
The results for the CRV3 model demonstrate
that a disk with $\Sigma_0 = 1440 \msol / {\rm pc^2}$
and $R_d = 5 \; \rm kpc$ 
is consistent, yet this implies that $> 95\%$ of the 
disk mass is in a component other than HI gas.
It is unlikely that $\Sigma_0$ approaches even 
$500 \msol / {\rm pc^2}$ given current estimates of the stellar 
(\cite{wol98}) and molecular (\cite{ge97}) baryonic fraction.
Furthermore, if $\Sigma_0 \approx 1500 \msol / {\rm pc^2}$ in
every damped system, this would imply a comoving density
approximately 100 times that of visible matter in the present
universe.
On the other hand if we include a massive dark halo ($v_h = 250 \mkms)$,
the resulting $\f{\delv}$ distribution (CRV4) is consistent
with the empirical distribution 
provided $R_h / R_d \lesssim 1$.  For larger values of $R_h/R_d$,
the rotation curve (CRV5)
rises too slowly to yield large $\delv$.
Finally, we investigate the effect of including a massive bulge in
models CRV6 and CRV7 where we adopt the same disk and halo curves
from models CRV4 and CRV5.
We find the presence of
a massive bulge $(v_b = 180 \mkms$) with $R_b \ll R_d$
has little impact on the $\f{\delv}$ distribution as the bulge
affects only the inner rotation curve.
To conclude, the TRD model requires a rotation curve which is nearly
flat for $R_d > 1$.  While this can be achieved with a massive disk 
resembling that of the Milky Way, it would requires a non-gaseous
baryonic component with comoving density two orders of magnitude 
greater than the density of visible matter in current galaxies, which
is implausible.
Assuming that the mass of the disk 
is dominated by the HI gas with 
$\Nperp \approx 10^{21} \cm{-2}$ requires the presence of a massive
dark halo with $R_h \lesssim R_d$.

\subsection{PHOTOIONIZATION}

Over the redshift range 
spanning our damped \lya observations the IGM is
primarily ionized, presumably by the ambient UV radiation field
from background quasars.
At redshift $z \sim 2.5$ the UV intensity is often
described as a modified power law (\cite{haa96})

\begin{equation}
J(\nu) \approx J_{912} \ltp {\nu \over \nu_{912}} \rtp^\alpha
\label{jnu}
\end{equation}

\noindent with 
$J_{912} \approx 10^{-21.5} \, {\rm erg/s \; 
s^{-1} \; cm^{-2} \; sr^{-1} \; Hz^{-1}}$ and exponent
$\alpha \approx -1$.
This background flux creates ionization fronts in the HI disks comprising
the TRD Model, analogous to those observed
in present spiral galaxies (\cite{mlny93}).   
Because a complete treatment involving
a solution of the radiative transfer equation is beyond the scope
of this paper, we address the problem with two different
approximations.  
In one case, we assume a sharp HI edge to the disks of the TRD model.
In the other, we assume the gas is photoionized if its volume density
is below some critical value.
The photoionization of the disk will eliminate from the derived statistical
sample those sightlines with $\N{HI} \approx N_{thresh}$.
Since these sightlines tend to penetrate the disk at large impact
parameters which yield small $\delv$, we predict a general shift in the 
$\f{\delv}$ distribution to
larger velocity widths.  By analogy to the results from warped disk models,
we expect the photoionized disk to mimic thicker disks.  

First, we consider the HI Edge Model,
which incorporates a sharp ``HI edge'' for the exponential disk
at a column density, $N_{ph}$. 
Using a 2D analogy to the Str$\rm \ddot o$mgren sphere argument, 
we find that the column density of ionized gas above the HI gas is
given by 

\begin{equation}
N_{ph} = {2 \phi \over \alpha_{rec} <n>} \cmma
\end{equation}

\noindent where $\phi$ is the flux of ionizing photons 

\noindent (i.e.\
$\phi = \intl_{\nu_H}^\infty d\nu \, 2 \pi J_\nu / h \nu$),
$<n>$ is the average hydrogen
volume density, and $\alpha_{rec}$ is the Case B recombination coefficient.
Given the above value for $J_{912}$, 
with $<n> \, = 1 \cm{-3}$ we find $N_{ph} \approx 10^{18} \cm{-2}$.
To simulate photoionization of the HI disks in the TRD model,
we systematically remove $N_{ph}$ of column normal to the
disk everywhere.  This leads to a sharp
HI edge at $R_{ph} = \ln [\Nperp / 2 N_{ph} ]$ and effectively
removes $2 N_{ph}$ from each sightline which penetrates
within $R_{ph}$.  
As $J_{912}$ is poorly constrained observationally, we consider
$J_{912}$ values ranging from $10^{-21.0} - 10^{-22.5} \, {\rm erg \; 
s^{-1} \; cm^{-2} \; sr^{-1} \; Hz^{-1}}$, where the 
favored value is $10^{-21.5}$ (\cite{haa96}).  To consider
meaningful $N_{ph}$ values, 
we relate $N_{ph}$ to $\Nperp$, $h$, and $J_{912}$ with the
following equation:

\begin{equation}
N_{ph} = 10^{19} \cm{-2} \ltp {h/R_d \over 0.2} \rtp
\ltp {J_{912} \over 10^{-21.5} } \rtp
\ltp {10^{21.2} \over \Nperp } \rtp  \perd
\end{equation}

\noindent The other approximation we make 
to describe the photoionization of the 
disk is to assume the gas is predominantly ionized below
a critical volume density, $n_{ph}$.
For the TRD model, the critical density can be defined as

\begin{equation}
n_{ph}  = n(R= R_{ph}, Z = 0) = {N_{ph} \over h}
\end{equation}

\noindent with $R_{ph}$ and $N_{ph}$ defined as above.
With $h = 0.3 R_d$, $R_d = 5$ kpc, and 
$N_{ph} = 10^{18} - 10^{20} \cm{-2}$ this implies density
cutoffs of $n_{ph} \approx [10^{-4} - 10^{-2} \cm{-3}]$.
We refer to this photoionization model as the Critical Density
model. 

\begin{figure}[ht]
\begin{center}
\includegraphics[height=5.0in, width=3.5in, bb = 65 58 547 734]{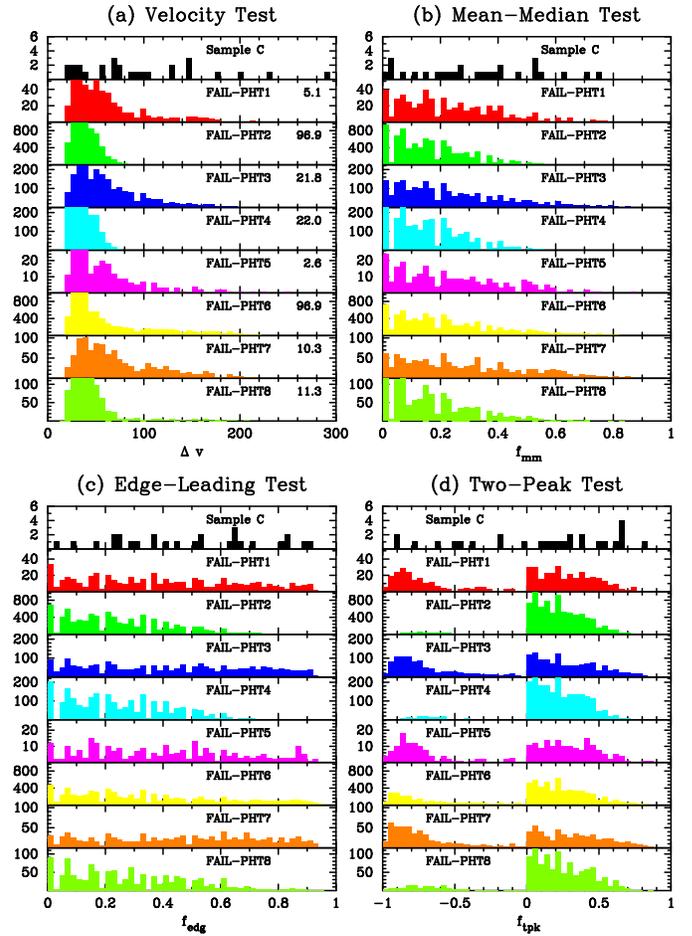}
\caption{Test statistic distributions for the {\it removed}
sightlines from four representative runs of the two photoionization
models considered in the text.
\label{elimfig}}
\end{center}
\end{figure}

\begin{table} \footnotesize
\caption{\label{Nphrslts}}
\begin{center}
{\sc Photoionization Models \smallskip}
\begin{tabular}{lccccc}
\tableline
\tableline \tskip
Label & Model\tablenotemark{a} &  
$h/R_d$ & 
$\log[\N{HI}]$ &
$-\log[J_{912}]$ &
$\log[N_{ph}]$ \\
\tableline \tskip
PHT 1 & HI\tablenotemark{b} & 0.3 & 21.2 & 22.0 & 18.7 \\
PHT 2 & HI & 0.3 & 21.2 & 20.5 & 20.2 \\
PHT 3 & HI & 0.3 & 20.8 & 21.5 & 19.6 \\
PHT 4 & HI & 0.1 & 21.2 & 21.0 & 19.2 \\
PHT 5 & CD\tablenotemark{c} & 0.3 & 21.2 & 22.0 & 18.7 \\
PHT 6 & CD & 0.3 & 21.2 & 20.5 & 20.2 \\
PHT 7 & CD & 0.3 & 20.8 & 21.5 & 19.6 \\
PHT 8 & CD & 0.1 & 21.2 & 21.0 & 19.2 \\
\tskip \tableline
\end{tabular}
\end{center}
\tablenotetext{a}{All models assume $v_{rot} = 225 \mkms$}
\tablenotetext{b}{HI Edge Model}
\tablenotetext{c}{Critical Density Model}
\end{table}

It is revealing
to examine the kinematic
characteristics of those sightlines {\it removed}
from the TRD statistical sample 
on account of photoionization. 
Figure~\ref{elimfig} plots the test statistics for the 
{\it eliminated} sightlines from 4 representative runs of
each photoionization model the parameters of which are given in
Table~\ref{Nphrslts}.  
Included in each panel of the $\f{\delv}$ distribution is the
percentage of sightlines eliminated.
Note that very few sightlines are removed in a 10000 sightline run for
$N_{ph} \approx 10^{18} \cm{-2}$, hence the effects of photoionization 
are minimal for small $J_{912}$, larger $\Nperp$ or small $h/R_d$.  
By contrast, simulations with $N_{ph} = 10^{20} \cm{-2}$ have very 
many sightlines removed so that there is a large impact on the
simulations. 
As expected, the vast majority of sightlines eliminated have small
$\delv$ and correspondingly small $\f{mm}$ and $\f{2pk}$ values.
Also, the results for thin disks are essentially unaffected
by photoionization as most sightlines yield small $\delv$ irrespective
of photoionization.
The results are very similar for the two photoionization
models, although the Critical Density model does remove 
a few more large $\delv$.  

\begin{figure}
\includegraphics[height=5.0in, width=3.5in, bb = 45 38 567 754]{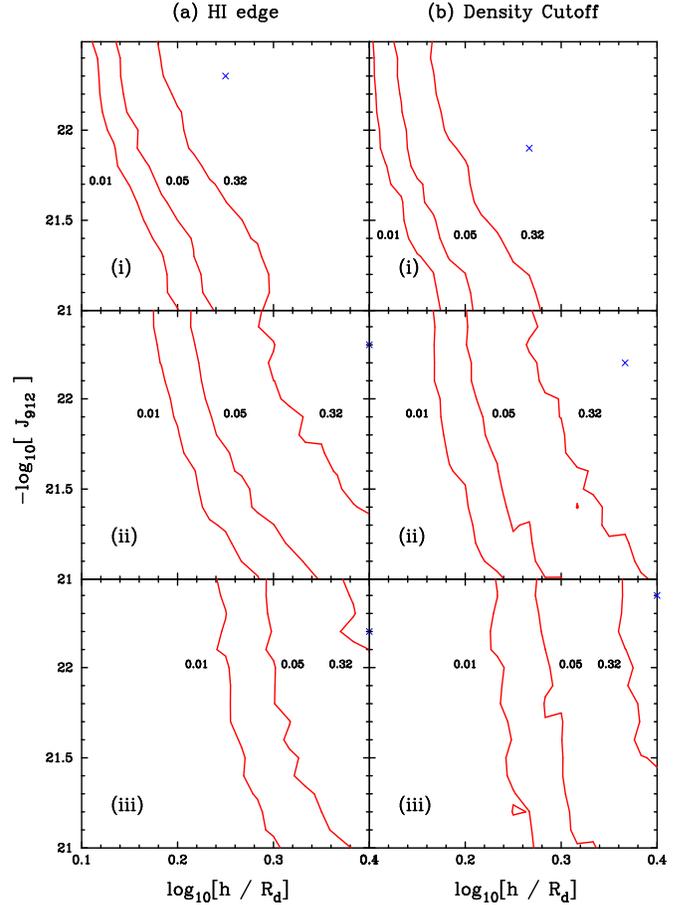}
\caption{Isoprobability contours for disks with 
(a) $\Nperp = 10^{20.8} \cm{-2}$, (b) $\Nperp = 10^{21.2} \cm{-2}$,
and (c) $\Nperp = 10^{21.6} \cm{-2}$, for a range of $h/R_d$ and
$J_{912}$ values in the HI Edge Model.
\label{hJks}}
\end{figure}

\begin{figure*}[ht]
\begin{center}
\includegraphics[height=5.0in, width=4.0in, angle=-90, bb = 45 38 567 754]{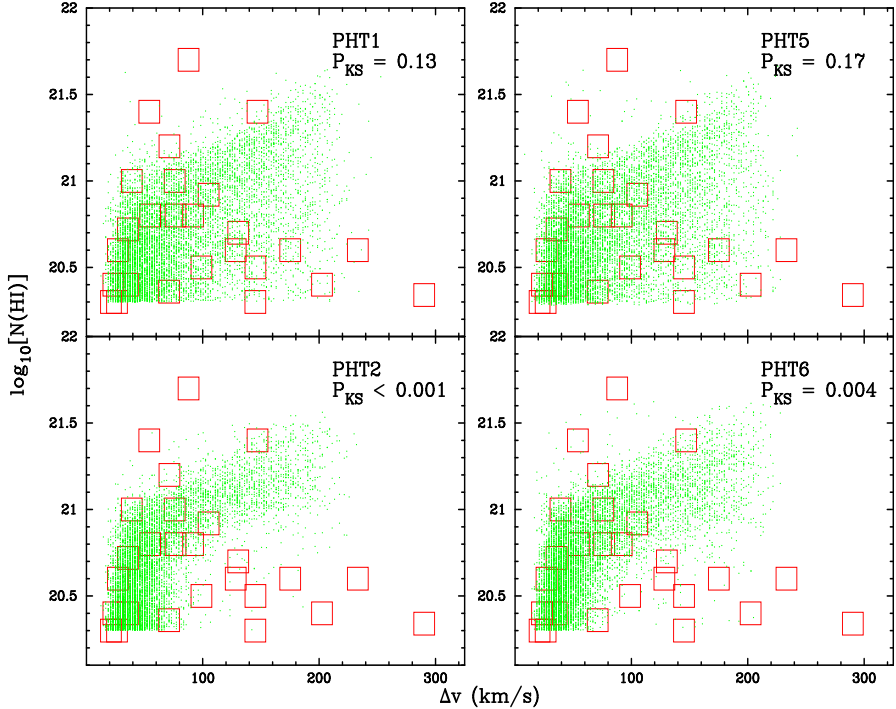}
\caption{
Plots of the $\N{HI}$ vs.\ $\delv$ pairs for the damped
\lya systems (boxes) and four photoionization models: 
PHT1 and PHT5 where $J_{912}$ is small and PHT2
and PHT6 where photoionization is significant.  
The $P_{KS}$ are derived from the 2-dimensional KS
Test.  Note that
photoionization clearly worsens the agreement between the
model and the empirical distribution.
\label{phNvks2}}
\end{center}
\end{figure*}

The results presented in Figure~\ref{elimfig} suggest
photoionization will improve the agreement between models with
small thickness or low rotation speed
(i.e.\ models with an $\f{\delv}$ distribution 
dominated by small $\delv$) and the empirical data set.  
However, besides eliminating sightlines with small $\N{HI}$,
photoionization of the edges of the disks tends to lower the average $\delv$
for a given sightline;
because of photoionization, the HI path along a given sightline is
reduced, which reduces the differential
velocity along the line of sight and on average lowers $\delv$.
The net result is that 
photoionization actually worsens the agreement for disks of all
thickness and central column density, contrary to our initial expectation. 
Figure~\ref{hJks} presents
isoprobability $P_{KS} (\delv)$ contours for disks with
(i) $\Nperp = 10^{20.8} \cm{-2}$, (ii) $\Nperp = 10^{21.2} \cm{-2}$,
and (iii) $\Nperp = 10^{21.6} \cm{-2}$, for a range of $h/R_d$ and
$J_{912}$ values in the (a) HI Edge model and (b) Critical Density model.  
The three contours correspond
to $P_{KS} (\delv)$ values of 0.01, 0.05, and 0.32 
and the $\times$ marks the highest $P_{KS} (\delv)$ value in the
explored parameter space.  The 
agreement clearly worsens for increasing
$J_{912}$ and decreasing $h/R_d$.  
If the UV background does have $J_{912} \approx 10^{-21}$, then
photoionization will have a significant impact on the results
derived for in the TRD Model.  
In particular, the thickness would have to exceed $h/R_d > 0.2$ 
for disks with $\Nperp = 10^{20.8} \cm{-2}$ and 0.3 for
disks with larger $\Nperp$.
While such large values for $J_{912}$ are unlikely, they are not
ruled out by current observations.  For the 
$J_{912} \approx 10^{-21.5}$ the effect on the TRD model is moderate, generally 
requiring less than a $30\%$ increase in the thickness of the disks.

Lastly, we examine how the photoionization models affect the
distribution of $\N{HI}$, $\delv$ pairs.  In PW (Figure 13)
we noted the
rather poor concordance between the TRD model and the damped \lya 
observations and suggested photoionization may improve the agreement.
Specifically, we expected photoionization to reduce the number of
sightlines with small $\delv$ and small $\N{HI}$ and
thereby improve the agreement with the damped \lya observations.
Figure~\ref{phNvks2} plots
the $\N{HI}$, $\delv$ pairs from Sample C (big squares)
and the $\N{HI}$, $\delv$ pairs for 4 of the photoionization models.
The PHT1 and PHT5 models indicate the results without significant
photoionization while the PHT2 and PHT6 models highlight its effects.
The $P_{KS}$ values were derived from the 2-dimensional 
Kolmogorov-Smirnov test (\cite{prss92}).
For the reasons discussed above, photoionization 
worsens the agreement between the TRD model and damped distributions. 
While models PHT2 and PHT6 are inconsistent at the 99$\%$ c.l.,
it should be noted, however, that one can improve the concordance 
by considering a range of $v_{rot}$ and $\Nperp$ values.
Also, as noted in Wolfe \& Prochaska (1998), the presence of many large
$\delv$, small $\N{HI}$ pairs may indicate the presence of
a 'hole' in the inner region of the disk as observed in many local
spirals.

\section{SUMMARY AND CONCLUSIONS}

We have presented new observations on the low-ion kinematics
of the damped \lya systems.  The full sample of 31
profiles confirms the primary conclusions of PW, in particular,
(i) models with kinematics dominated by random 
or symmetric velocity fields are inconsistent with the damped \lya
kinematics, (ii) the TRD model, which consists of a population of
thick, rapidly rotating disks at high $z$, naturally reproduces
both the observed edge-leading asymmetry of the empirical profiles
as well as the distribution of velocity widths, and (iii)
models incorporating centrifugally supported disks within the framework
of the standard CDM cosmology are ruled out at high levels of confidence.
In addition, a comparison of the kinematic properties of profiles
of the highest redshift systems ($\bar z = 3.24$) with the lower 
redshift systems ($\bar z = 2.06$) reveals no significant evolution
in the kinematics of the damped \lya systems.  This last
observation may
place strong constraints on scenarios of galaxy formation which
predict significant evolution over this epoch.

Presently there are two working models which explain the kinematic
characteristics of the damped \lya systems:  (1) the TRD model
and (2)  merging protogalactic clumps in
numerical simulations of the standard Cold Dark Matter 
cosmology (\cite{hae97}).
In this paper we have focused on the TRD model.  In particular,
we have investigated the robustness of the model to including
more realistic disk properties, specifically
disk warping, physical rotation curves and 
photoionization.  
Given the prevalence of warping in local disk galaxies, we
considered its effects on the kinematics of the disks in the
TRD model.
We found that the results of warping are dominated
by two competing effects.  Sightlines which penetrate
both the unwarped inner disk and the warped outer disk yield
moderately higher $\delv$ than those simply intersecting an
unwarped disk.  At the same time, however, some warped disks
have significantly larger cross-section to sightlines with 
large impact parameters which tend to yield small $\delv$.
Having considered a number of warped disks with a broad range
of properties, we find:
(i) in extreme cases, warping mimics disks with up to
$50\%$ larger or smaller effective thickness ($h/R_d$ value),
(ii) warping leads to very few extra large $\delv$ values in the 
$\f{\delv}$ distribution and
therefore has little consequence on the acceptable values for $v_{rot}$, 
and (iii) the lower limit to $h/R_d$ is nearly unchanged as we find
$h$ must be $> 0.1 R_d$ for both warped and unwarped disks.

In PW, we assumed a flat rotation curve, $v_\phi = v_{rot}$,
extending from $R = 0 \to \infty$ and $Z = 0 \to \infty$.
In this paper we adopted rotation curves derived from specific bulge,
halo, and disk components.
Assuming an exponential profile
is a good description of the density profile for the damped \lya
systems, we find the rotation curves derived from gravity 
generated by the HI gas
{\it alone} cannot reproduce the empirical $\f{\delv}$ distribution.
If the rotation curve is dominated by the disk, 
one must introduce another mass component (e.g.\ stars, molecules)
to establish consistency.
At the same time we find that the rotation curve
derived from a massive
halo with core radius $R_h \lesssim R_d$ also yields a
$\f{\delv}$ distribution consistent with the observations.
We believe that this latter explanation is more plausible.
We also find the presence of the bulge to be largely inconsequential.

Lastly, we studied the effects of the intergalactic photoionizing
background radiation
 on the disk kinematics.  We made two separate approximations
to model the photoionization of the disks: (a) an HI edge model where
the disk is photoionized at radii $R > R_{ph}$ with $R_{ph}$ set by
the intensity of the photoionizing background and the disk properties
and (b) a Critical Density model where all gas with volume density
$n \leq n_{ph}$ is presumed ionized.  Contrary to our expectations,
we find that photoionization tends to worsen the agreement between
the TRD model and the damped \lya observations.  The effect, however,
is not large ($ < 30 \%$) for the favored value of 
$J_{912} = 10^{-21.5}$, but for $J_{912} = 10^{-21}$ a substantial
$(> 50 \%)$ increase in the thickness of the disks would be required.

In summation, then, we find the TRD model is very robust to 
tests against the damped \lya observations.  
The challenge remains, however, to consistently
incorporate this model within a cosmological framework.
While the clump model fits naturally within the SCDM cosmology,
it must be demonstrated that
the clump model exhibits similar robustness to comparisons
with the damped \lya kinematics.
While Haehnelt et al.\ (1997) did show that the clump model could
explain the damped \lya observations from PW for a single set
of parameters, a formal investigation of the full physical parameter
space with meaningful statistics
has yet to be performed.  In addition, it is not clear
how that model will change given different cosmological parameters,
e.g.\ an Open Universe where merging plays a smaller role
at $z \approx 2.5$.  The model must also be tested against the
new observations, in particular the new $\f{\delv}$ distribution 
which extends to $\approx 300 \mkms$.  
Finally, the fact that the numerical simulations do not reproduce the 
observed properties of modern
galaxies when evolved to the present universe (\cite{nav95})
suggests the model may have serious inconsistencies in the
early universe.

In future papers we will introduce observations of the high-ion
transitions (e.g.\ CIV~1548) with the aim of 
further constraining
the two working models as well as advancing our understanding of
the ionized gas associated with the damped \lya systems.
This gas is
presumed to reside in the halo of these protogalaxies and therefore
may give more direct indications of the dark matter associated with
the damped \lya systems.
We also intend to consider effects (e.g.\ multiple disks) which
would improve the agreement between the semi-analytic models
of standard cosmology (\cite{kau96,mmw97}).

\acknowledgements

We thank the group headed by W. L. W. Sargent including Limin Lu
for generously providing us with their HIRES spectra. 
We also would like to thank R. Becker, S. Burles,
L. Storrie-Lombardi and I. Hook
for providing targets.
The authors would also
like to thank Tom Barlow for his excellent HIRES data
reduction software.  Finally, we wish to acknowledge many
helpful comments by the anonymous referee.
AMW and JXP were partially supported by 
NASA grant NAGW-2119 and NSF grant AST 86-9420443.


\begin{thebibliography}{}

\bibitem[Becker 1998]{bin98}
Becker, R.H., private communication

\bibitem[Binney and Tremaine 1987]{bin87}
Binney, J. and Tremaine, S. 1987, {\it Galactic Dynamics}
(Princeton: Princeton University Press), p. 229

\bibitem[Briggs 1990]{brg90}
Briggs, F.H. 1990, \apj, 352, 15

\bibitem[Casertano 1983]{cast83}
Casertano, S. 1983, \mnras, 203, 735

\bibitem[Eggen, Lynden-Bell, \& Sandage 1962]{egg62}
Eggen, O.J., Lynden-Bell, D. \& Sandage, A. 1962,
\apj, 136, 748

\bibitem[Ge \& Bechtold 1997]{ge97}
Ge, J. \& Bechtold, J. 1997, \apj, 477, 73

\bibitem[Gradshteyn \& Ryzhik 1980]{grd80}
Gradshteyn, I.S., \& Ryzhik, I.M. 1980. {\it Tables of 
Integrals, Series and Products.} New York: Academic

\bibitem[Haardt \& Madau 1996]{haa96}
Haardt, F. \& Madau, P. 1996, \apj, 461, 20

\bibitem[Haehnelt et al.\ 1997]{hae97}
Haehnelt, M.G., Steinmetz, M. \& Rauch, M. 1997, astro-ph 9706201

\bibitem[Hernquist 1990]{hern90}
Hernquist, L. 1990, \apj, 356, 359

\bibitem[Jedamzik \& Prochaska 1998]{jdzk98} 
Jedamzik, K. \& Prochaska, J. X. 1998,
{\it \mnras}, in press  (astro-ph 9706290)

\bibitem[Kauffmann 1996]{kau96}
Kauffmann, G. 1996, \mnras, 281, 475

\bibitem[Maloney 1993]{mlny93}
Maloney, P. 1993, \apj, 414, 41

\bibitem[Mo, Mao, \& White 1997]{mmw97}
Mo, H.J., Mao, S., \& White, S.D.M. 1997, astro-ph/9707093

\bibitem[Navarro, Frenk \& White 1995]{nav95}
Navarro, J.F., Frenk, C.S., \& White, S.D.M. 1995,
\mnras, 275, 56

\bibitem[Press 1992]{prss92}
Press, W. H. 1992, Numerical Recipes in FORTRAN,
(New York: Cambridge University Press)

\bibitem[Prochaska \& Wolfe 1997b]{pro97b}
Prochaska, J. X. \& Wolfe, A. M. 1997, \apj, 486, 73

\bibitem[Sancisi 1998]{sanc98}
Sancisi, R.  1998, private comm.

\bibitem[Wolfe et al.\ 1995]{wol95a}
Wolfe, A. M., Lanzetta, K. M., Foltz, C. B., and
Chaffee, F. H. 1995, \apj, 454, 698

\bibitem[Wolfe 1995]{wol95b}
Wolfe, A.M. 1995, in {\it ESO Workshop on
QSO Absorption Lines}, ed. G. Meylan, (Berlin:Springer-Verlag), p. 13

\bibitem[Wolfe $\&$ Prochaska 1998]{wol98}
Wolfe, A.M. $\&$  Prochaska, J. X., \apj, 494, 15L

\end{thebibliography}
\end{document}